\documentclass[a4paper,onecolumn,superscriptaddress,10pt]{article}
\pdfoutput=1

\usepackage{authblk}                       % for author/affiliation

\usepackage[utf8]{inputenc}
\usepackage[english]{babel}
\usepackage[T1]{fontenc}
\usepackage{amsmath}
\usepackage{hyperref}
\usepackage{tikz}
\usepackage{lipsum}
\usepackage{subcaption}
\usepackage{amsmath}
\usepackage{amssymb}
\usepackage{amsfonts}
\usepackage{amsthm}
\newcommand{\PathAttempt}{\mathrm{PathAttempt}}
\newcommand{\OpenHorn}{\mathrm{OpenHorn}}
\newcommand{\U}{\mathcal{U}}
\newcommand{\Adm}{\mathsf{Adm}}
\newcommand{\Face}{\mathsf{Face}}
\newcommand{\Simplex}{\mathsf{Simplex}}
\newcommand{\Enc}{\mathsf{Enc}}
\newcommand{\Embed}{\mathsf{Embed}}
\newcommand{\ET}{\mathsf{ET}}
\newcommand{\Token}{\mathsf{Token}}
\newcommand{\Label}{\mathsf{Label}}

\newcommand{\RepBar}{\mathsf{RepBar}}
\newcommand{\op}{\mathsf{op}}
\newcommand{\Base}{\mathsf{Base}}
\usepackage{stmaryrd}
\usepackage{mathtools}
\usepackage{bm}
\newcommand{\Corpus}{\mathcal{C}}
\usepackage{geometry}
\usepackage{microtype}
\usepackage{booktabs}
\usepackage{mathpartir}
\usepackage{tikz}
\usetikzlibrary{cd,arrows,calc}
\newcommand{\NoCarryYet}{\mathsf{NoCarryYet}}

\newcommand{\RupLed}{\mathsf{RupLed}}

\newcommand{\DynSem}{\mathbf{DynSem}}

\newcommand{\Rupture}{\mathsf{Rupture}}

\newcommand{\transport}{\mathsf{transport}}

\providecommand{\SSet}{\mathbf{sSet}}

\newcommand{\Time}{\mathbb{T}}

\newcommand{\coloneqq}{\mathrel{\mathop:}=}

\newcommand{\Id}{\mathsf{Id}}

\newcommand{\refl}{\mathsf{snd}}
\newcommand{\ev}{ev}
\newcommand{\Carry}{\mathsf{Carry}}
\newcommand{\ExposeGlue}{\mathsf{ExposeGlue}}
\newcommand{\Ex}{\mathrm{Ex}}
\newcommand{\ExInf}{\Ex^{\infty}}
\theoremstyle{plain}
\newtheorem{theorem}{Theorem}[section]
\newtheorem{lemma}[theorem]{Lemma}
\newtheorem{corollary}[theorem]{Corollary}
\theoremstyle{definition}

\newtheorem{example}[theorem]{Example}
\theoremstyle{remark}
\newtheorem*{remark}{Remark}
\newtheorem*{proposition}{Proposition}

\newcommand{\Edge}{\mathsf{Edge}}
\begin{document}

\title{\textbf{DHoTT: A Temporal Extension of Homotopy Type Theory for Semantic Drift}}
\author{
  Iman Poernomo
}
\affil{
  Data and Analytics, London Stock Exchange Group\\
  \texttt{iman.poernomo@gmail.com}\\
  \url{http://substack/imanpoernomo}
}
\maketitle

\begin{abstract}
\noindent
Dynamic HoTT (DHoTT) is a conservative extension of Homotopy Type Theory tailored to
\emph{evolving texts}—long-form prompt/response histories in conversational AI.
In a typical chat system, a large language model (LLM) is repeatedly queried with a
growing \emph{prefix}. At turn $\tau$ the input is the concatenation of all previous
prompts and responses $C(\tau)$, and the resulting reply is appended to form
$C(\tau{+}1)$.  We study the logical semantics of these time-indexed texts and how
their meanings drift or break over time. We do this by relating Homotopy Type Theory to
distributional semantics and to topological data analysis (TDA) on embedding spaces.
The novelty of our work lies in this synergy and in treating conversational time as a
chain of slices $\tau \mapsto C(\tau)$.

Our starting point is a measurement-to-reasoning discipline grounded in standard
embedding practice.  We take an evolving text $C(\tau)$ arising from a human--LLM
conversation. For each turn $\tau$ we consider all occurrences of word tokens seen
so far.  Using a frozen encoder (for example, the penultimate layer of an LLM), we map these
tokens to the unit sphere $S^{d-1}$. We then build a good cover by spherical caps and form
the Čech nerve.  Passing to a Kan fibrant replacement yields a Kan complex
$ET(\tau)$—the \emph{Evolving Text} at time $\tau$. In this complex, identity types are path
spaces and dependent types admit ordinary HoTT transport.  Time itself is modelled
presheaf-wise by a functor $ET:\Time^{\mathrm{op}}\to\SSet$. All HoTT rules
interpret fibrewise, giving familiar substitution, fibrancy, and conservativity
properties.

On top of this we introduce a small cross-time calculus for semantic change.  We add
two proof-relevant primitives: \emph{carry}, which records when a later use has a
certified path back into an earlier fibre under a given admissibility policy, and
\emph{rupture}, a positive $\Sigma$-type that stores finite, policy-checked failed
attempts together with an open-horn tag.  An append-only \emph{ledger} accumulates
these witnesses so that earlier ruptures are not erased when later carries appear.
This yields a minimal, compositional kernel for talking about ``semantic drift'' over
conversational time. The result is both logically typed and geometrically grounded.

We illustrate the construction on small, replicable token/sentence examples derived
from real conversational data, using a concrete DeBERTa-based embedding pipeline.  Our
results provide a foundational bridge between contextual/distributional semantics,
topological data analysis on embedding spaces, and HoTT-style semantics. We argue
that they offer a principled core for analysing and auditing the dynamics of large
language models and other embedding-based systems.
\end{abstract}

\section{Introduction}\label{sec:intro}

\paragraph{Problem.}
Consider any conversation between two friends on a Saturday afternoon.  They sit on
the river bank, chatting about the current. One suggests stopping by the bank to take
out cash. Later they part ways at Bank station in London.  Over the course of a
single afternoon, the word \emph{bank} has wandered through at least three distinct regions
of use.  The same phenomenon occurs with other words (\emph{freedom}, \emph{cat}) and over
longer time scales. Conversational meaning typically shifts. The same surface form
is repeatedly reused in subtly or radically different ways as a dialogue unfolds.

Existing work offers many perspectives on this problem.  Dynamic semantics in
linguistics treats meaning as context update rather than static truth conditions.
Distributional and contextual semantics model ``meaning-as-use'' via embeddings and
similarity in vector space. In this paradigm, the meaning of a word is characterized by
the contexts in which it appears, encoded as vectors in a high-dimensional space.
Topological data analysis (TDA) studies the shape of such
high-dimensional embedding spaces. This includes the cluster and loop structure of usage
patterns.  Each of these perspectives provides tools for \emph{measuring} how uses
cluster or diverge. But none by itself gives a typed, compositional language for
talking about coherence and incoherence of meaning. We need to reason both \emph{within} a conversational
turn (across tokens in a slice) and \emph{between} turns (across slices).

What we lack is a \emph{logic} that accounts not only for \emph{when} such shifts occur,
but also for \emph{what} constitutes lawful continuation, failure, and repair of meaning
across steps. This logic must be grounded in \emph{embedding-based} models rather than an
abstract toy semantics.

Our goal here is to show how a homotopical type theory, combined with simple
constructions from TDA, can make textual and cross-time reasoning \emph{proof-relevant}.
In particular, we focus on the setting where one of the friends in the conversation is an LLM. Embeddings come from a
contextual encoder. It is therefore particularly cogent to ask: how can we track, \emph{inside} a type theory, when a
model's conversation carries a sign coherently through its various ``banks'', and when it
fails or hallucinates?

The construction is straightforward. Given a finite cloud of contextual embeddings in a metric space, standard TDA machinery—good covers, Čech nerves, and the Nerve Theorem—produces a simplicial complex that captures the measured geometry. The Kan fibrant replacement $\ExInf$ then yields a space where identity types are path spaces and transport has the usual HoTT meaning. This is the same pipeline used in persistent homology and topological data analysis~\cite{Carlsson2009TDA,EdelsbrunnerHarer}, now applied to embedding spaces from language models. The novelty lies not in the mathematics but in recognizing that this standard construction grounds homotopy-theoretic reasoning about meaning in empirical, measurable data.

\paragraph{Motivation from conversational AI.}
Our primary motivation comes from the baseline architecture of modern conversational AI
systems.  A typical chat engine maintains a growing dialogue history. It repeatedly
queries a large language model (LLM) with the entire prefix.  At conversational turn
$\tau$ we write the current text as
\[
  C(\tau) = [\,s_1,\dots,s_{n_\tau}\,],
\]
where $s_i$ are utterances (user prompts or model responses).  The next turn appends a
new prompt/response pair to this sequence, yielding $C(\tau+1)$. The process
continues.  We call the family $\{C(\tau)\}_\tau$ an \emph{evolving text}: a text that
grows turn by turn as the conversation unfolds.  In practice, the model is always
conditioned on the latest $C(\tau)$. We would like to understand how the meanings
of individual words and phrases evolve as $\tau$ advances.

A natural way to interrogate an LLM's ``sense'' of each token at time $\tau$ is to read
out hidden states from an internal layer of a frozen encoder.  Given $C(\tau)$, we feed
it through a contextual encoder. Examples include a penultimate transformer block of a DeBERTa or
BERT-style model. For each occurrence $t$ in context $C(\tau)$ we obtain a
\emph{contextual embedding} $e_{\tau,t}\in\mathbb{R}^d$ that depends on the entire
history seen so far.  After normalising to the unit sphere, these embeddings give a
finite cloud of points in a high-dimensional metric space. This is a kind of instantaneous
``palpation'' of how the model currently represents the meanings of the signs in play at
time $\tau$.  The question is then: how can we reason \emph{within} such a slice about
semantic neighbourhoods and relations? And how can we reason \emph{across time} about
how those meanings drift or rupture as the conversation proceeds?

At a high level, our measurement layer follows a standard TDA recipe. For each turn $\tau$, we
treat the contextual embeddings for $C(\tau)$ as a finite sample in a metric space on
the unit sphere. We summarise that cloud by overlapping semantic regions (spherical caps).
We then form the Čech nerve of this cover.  The nerve is a simplicial complex that records
which usages cohabit which regions. A Kan fibrant replacement of this nerve then yields
a Kan complex $ET(\tau)$ equipped with a robust path calculus.  Intuitively, $ET(\tau)$
is ``the shape of use'' of the model at time $\tau$. Its points are contextualised
occurrences. Its $1$–simplices are semantic adjacencies. Its higher simplices
encode higher-order overlaps.  Section~\ref{sec:single-slice} makes this construction
precise. We describe embeddings on $S^{d-1}$, good covers (families of convex regions whose intersections are contractible; see Lemma~\ref{lem:good-cover-caps}), Čech versus Vietoris–Rips, and the
$Ex^\infty$ replacement. For the introduction it suffices to view $ET(\tau)$ as a
single-time Kan space in which identity types are path spaces and dependent types admit
ordinary HoTT transport.

To speak \emph{across} time, we view the family $\tau \mapsto ET(\tau)$ as a simplicial
presheaf on the poset of times.  Formally, we take $ET:\Time^{\mathrm{op}}\to\SSet$ and
reason fibrewise. Evaluation at $\tau$ preserves the Kan–Quillen structure. The
slice over the representable $y(\tau)$ is equivalent to presheaves on
$(\Time/\tau)^{\mathrm{op}}$, not to $\mathbf{SSet}$ itself.  All constructions that
matter for soundness—transport, the $\Sigma$-types used for rupture, and any standard
higher inductive types we adopt—are justified pointwise in the Kan fibres $ET(\tau)$.
On top of this measured semantics we place a small cross-time calculus, Dynamic HoTT
(DHoTT). This calculus conservatively extends HoTT with two proof-relevant primitives:
\emph{carry} and \emph{rupture}, plus an
append-only \emph{ledger} that stores these witnesses over time.

\begin{remark}[Carry and Rupture at a glance]
Before we develop the formal machinery, here is an informal summary of the two key primitives:
\begin{itemize}
  \item \textbf{Carry} records when a later occurrence of a sign has a certified path
    back into an earlier fibre under a chosen admissibility policy. This path serves as a
    receipt that licenses transport of dependent data.
  \item \textbf{Rupture} records finite, policy-checked failures. When no admissible witness exists,
    we store \emph{why} attempts fail as a structured $\Sigma$-type in the earlier fibre,
    together with an open-horn tag.
  \item \textbf{Ledger-based healing} occurs when a later seam of text re-embeds a sign
    so that the same policy now yields a carry. We append a success entry
    to an append-only ledger. Healing is not a new primitive—it is
    the observable pattern that failed attempts eventually give way to a
    carry with a canonical path in $ET(\tau)$.
\end{itemize}
These are ordinary HoTT paths (or their constructive absence) in $ET(\tau)$.
The novelty lies in providing proof-relevant provenance and an explicit ledger of
attempts to heal. This gives us a formal ``playback log'' for the journey of a sign over the evolution of its text.
\end{remark}

In this reading,
carries correspond to model-internal continuations of meaning. Ruptures expose
cases where the LLM fails to connect a later use to earlier structure.  Our examples in
Section~\ref{sec:example} are toy instances of this pattern on short conversational
snippets. The logic is designed to scale to richer dialogues and to support
downstream tools for drift and hallucination diagnostics.

\paragraph{Scope and limitations.}
We work with one concrete measurement pipeline
(embeddings $\to$ Čech cover $\to$ $\Ex^\infty$) because it is simple and
replicable. Other choices are possible. Different encoders, distances, complexes, or filtration
strategies may be better suited to particular domains of text. Examples include code development versus informal chatbot-style textual evolution. Our
results should be read as a \emph{template} for combining measured geometry with HoTT-style reasoning over any evolving text.

\paragraph{Contributions.}
Within that scope, this paper makes four contributions:

\begin{enumerate}
  \item \textbf{Fibrewise Kan semantics from measurements.} We present a concrete and
        replicable pipeline—embeddings $\to$ Čech nerve (good cover)
        $\to \Ex^\infty$. This yields $\ET(\tau)$. Identity and transport are
        those of HoTT in the Kan fibre. This aligns standard TDA foundations
        with simplicial semantics for embedding-based language models.

  \item \textbf{Cross-time calculus with receipts.} We provide a small, fibre-indexed
        extension of HoTT with formal rules for \emph{carry} and \emph{rupture}.
        These produce explicit witnesses (or explicit finite failures) in the
        earlier slice. We interpret this in $[\Time^{op},\mathbf{sSet}]$. The calculus
        is conservative over HoTT at each fixed $\tau$.

  \end{enumerate}

\paragraph{Roadmap.}
Section~\ref{sec:single-slice} develops the measured single-slice semantics on $S^{d-1}$. We describe embeddings from a frozen
encoder, Čech versus Vietoris–Rips complexes, the Nerve Theorem, and the passage
$N(U)\mapsto Ex^\infty N(U)$ to obtain the Kan base $ET(\tau)$. Section~\ref{sec:syntax-core} introduces the Dynamic
HoTT (DHoTT) calculus over $ET$, including the carry/rupture primitives, the ledger, and their
interpretation as a simplicial presheaf in $[\Time^{\mathrm{op}},\mathbf{sSet}]$. Section~\ref{sec:semantics} presents the simplicial-presheaf semantics and proves soundness, fibrancy, and conservativity. Section~\ref{sec:example} works through a
concrete conversational ``bank/cat/flow'' example. It shows how carries, ruptures, and ledger-based
healing appear in practice.  Section~\ref{sec:future} discusses implications for LLM analysis and prompt
engineering, situates DHoTT among related type-theoretic frameworks, and outlines AI-facing
applications and open problems.

\begin{remark}[Notation guide]
Key notation is introduced as follows:
$\ET(\tau)$ (Evolving Text base at time $\tau$) in Section~\ref{sec:single-slice};
$\Carry$, $\Rupture$, $\RupLed$ (cross-time primitives) in Section~\ref{sec:syntax-core};
$r_{\tau,\tau'}$ (restriction maps) in the assumption at the start of Section~\ref{sec:syntax-core};
$\Adm$ (admissibility policy) in Section~\ref{subsec:rupture}.
The judgement form $\Gamma \vdash_\tau J$ is read as ``judgement $J$ is provable at time $\tau$ in the fibre $\ET(\tau)$.''
\end{remark}

\section{Measured semantics for a text}
\label{sec:measured-single-text}\label{sec:single-slice}

This section fixes the geometric objects we use to measure a text. We work on the unit sphere
$S^{d-1}$ with the angular (great–circle) metric. We build a good cover by spherical caps of radius
$<\pi/2$. We take the Čech nerve. We then pass to a Kan fibrant replacement via $\ExInf$ for a
well–behaved path calculus. No temporal or type–theoretic machinery is used here yet. We work at a
single time~$\tau$ with a single snapshot of the text.

The construction follows standard topological data analysis. Good covers on Riemannian manifolds,
the Čech nerve construction, and the Nerve Theorem are classical results in algebraic
topology~\cite{goerss2009simplicial,EdelsbrunnerHarer}. Their application to point clouds in metric
spaces is well-established in persistent homology~\cite{Carlsson2009TDA}. What is new here is the
recognition that contextual embeddings from language models—points on the unit sphere $S^{d-1}$—admit
the same treatment, and that the resulting Kan complexes provide exactly the structure needed for
HoTT-style reasoning about distributional meaning.

\begin{remark}[Measurement precedes reasoning]
Everything in this section is \emph{measurement}. We extract embeddings, form covers, and build
nerves. In Section~\ref{sec:syntax-core} we will perform \emph{reasoning}—composing paths,
eliminating horns, and forming judgements. This strict separation ensures that we never
``measure our inferences.''
\end{remark}

\paragraph{Setup (embeddings on $S^{d-1}$).}
We assume a \emph{frozen} contextual encoder (one whose weights are not updated during our analysis)
\[
  \Enc : \textsf{Sequences} \longrightarrow \mathbb{R}^d
\]
such as BERT, DeBERTa, or a sentence encoder built on top of them
\cite{devlin2019bert,he2021deberta,reimers2019sentence}. We use a frozen encoder to ensure
that the measured geometry reflects the model's fixed internal representation, not training dynamics.
Given the text
$C(\tau) = [s_1,\dots,s_{n_\tau}]$ at time~$\tau$, we feed it through $\Enc$
and, for each token (or sentence) occurrence $t$ in context $C(\tau)$, we extract a
\emph{contextual embedding}
\[
  e_t \ :=\ \Enc_\ell(C(\tau),t) \in \mathbb{R}^d
\]
from a fixed internal layer~$\ell$. This is the standard notion of a contextual
embedding used in current NLP practice: $e_t$ depends on the entire sequence,
not just on the surface form of $t$.

We then normalise to the unit sphere by setting $\hat e_t := e_t/\|e_t\|_2 \in S^{d-1}$ and adopt
the angular (cosine) metric
\[
  d_\angle(u,v) := \arccos\langle u,v\rangle \in [0,\pi].
\]
The result is a finite metric sample $P \subset S^{d-1}$ of contextual embeddings for this
single time~$\tau$. All that follows depends only on $P$ as a finite subset of a metric sphere.
No model–specific facts about $\Enc$ or transformer architectures are used in the mathematics.

\paragraph{From points to regions (spherical caps and good covers).}
We summarise dense usage by (open) spherical caps on $S^{d-1}$:
\[
  B(\mu,\rho) := \{ x \in S^{d-1} \mid d_\angle(x,\mu) < \rho \}, \quad 0 < \rho < \pi/2.
\]
Caps of radius $< \pi/2$ are geodesically convex. Any two points in a cap have unique
minimising geodesics of length $<\pi$ lying in the cap. Since intersections of geodesically
convex sets are geodesically convex, every nonempty finite intersection of such caps is
contractible. Thus a family $U = \{B_j\}$ with $\rho_j < \pi/2$ forms a good cover of
$\bigcup_j B_j$.

\begin{lemma}[Geodesic caps form a good cover on $S^{d-1}$]\label{lem:good-cover-caps}
Let $U=\{B_j\}_{j\in J}$ be a locally finite family of geodesic caps on the round sphere
$S^{d-1}$ with angular metric, where
\[
  B_j \;=\; \{\,x\in S^{d-1}\mid d_\angle(x,\mu_j) < \rho_j\,\}
  \qquad\text{and}\qquad 0<\rho_j<\pi/2.
\]
Then every nonempty finite intersection $\bigcap_{r=0}^k B_{j_r}$ is geodesically convex and
hence contractible. In particular, $U$ is a good cover of $\bigcup_j B_j$.
\end{lemma}

\begin{proof}
On the round sphere, the \emph{convexity radius} at every point equals $\pi/2$.
For any $0<\rho<\pi/2$, the geodesic ball $B(\mu,\rho)$ is \emph{strongly geodesically convex}.
Any two points $x,y\in B(\mu,\rho)$ are joined by a unique minimizing geodesic of length $<\pi$
lying entirely inside $B(\mu,\rho)$ (see, e.g., \cite[§6.6]{LeeRiemannian} or
\cite[Prop.~5.6]{doCarmo}).

Each $B_j$ is such a ball, so it is strongly geodesically convex. Finite intersections of strongly
geodesically convex sets are again geodesically convex. The common minimizing geodesic remains in
every factor. Hence each nonempty $\bigcap_{r} B_{j_r}$ is geodesically convex. A geodesically
convex subset of a sphere is contractible by straightening along its unique minimizing geodesics
to a chosen basepoint. Therefore $U$ is a good cover.
\end{proof}

\paragraph{The Čech nerve and the Nerve Theorem.}
The Čech nerve $N(U)$ is the abstract simplicial complex with vertices $\{[j]\}$ and simplices
$[j_0,\dots,j_k]$ exactly when $\bigcap_r B_{j_r}\neq\varnothing$. 
The Nerve Theorem then yields a homotopy equivalence
$|N(U)|\simeq \bigcup_j B_j\subseteq S^{d-1}$ for good covers
(see, e.g., \cite{EdelsbrunnerHarer,goerss2009simplicial}).

We regard
$N(U)$ as the measured ``shape of use'' in this slice of the model's embedding space.

\paragraph{Token/incidence view (Dowker dual).}
Let $T$ be the set of token occurrences. We define the incidence relation
$R\subseteq T\times J$ by $(t,j)\in R \iff \hat e_t\in B_j$. Dowker's duality identifies the
homotopy type of the token complex $K_T(R)$ (simplices are token sets sharing some basin) with that
of the basin complex $K_J(R^\top)$ (simplices are basins jointly inhabited by a token). This is equivalent to
$N(U)$ up to subdivision. We use both views. Basins serve as coarse regions; tokens serve as fine witnesses.
In the LLM setting, $K_T(R)$ encodes which contextual embeddings produced by $\Enc$ cohabit
semantic basins at time~$\tau$.

\paragraph{Measurement vs.\ reasoning (strict split).}
\emph{Measurement} is performed on the raw Čech object $N(U)$. If desired, we perform measurement along a filtration
in $\rho$ for barcodes. \emph{Reasoning}—composing paths and eliminating horns—happens only after
a Kan fibrant replacement. We do not ``heal'' $N(U)$ by adding faces. Fillers appear only in the
reasoning object. This separation will be crucial later when we interpret logical judgements in a
Kan fibre while keeping the underlying measurements fixed.

\paragraph{Kan fibrant replacement (path calculus without changing homotopy type).}
Raw nerves typically contain open horns. We therefore set
\[
  A \ :=\ \ExInf\,N(U)\ \in\ \SSet_{\mathrm{Kan}}.
\]
The unit $\eta_\infty:N(U)\to A$ is a weak equivalence in $\SSet$, so homotopy type is preserved.
Every horn in $A$ has a filler and path composition is coherently associative. In pictures,
we may mark measured Čech adjacencies as \emph{solid} and $\ExInf$–licensed short composites as
\emph{dashed}. Only the latter live in $A$. In the notation of the introduction, $A$ is the
single-slice version of $\ET(\tau)$.

\paragraph{Why Čech (and a note on Vietoris–Rips).}
Čech records \emph{joint} presence via multi–overlaps. Vietoris–Rips (VR) is often used as a
computational proxy built from pairwise distances. There are standard inclusions/interleavings
between Čech and VR complexes~\cite{EdelsbrunnerHarer}. For example, $\check C(r)\subseteq
\mathrm{VR}(r)\subseteq \check C(\sqrt{2}\,r)$ in Euclidean settings. On the sphere with angular
metric and caps of radius $<\pi/2$, the same qualitative behaviour holds locally. Nevertheless VR
may over–fill loops using only pairwise edges in higher dimensions. We therefore retain Čech as the
semantic ground truth and use VR only as a computational proxy when needed.

The choice of angular metric and spherical caps is dictated by the data, not by mathematical convenience.
Cosine similarity—equivalently, angular distance on the unit sphere—is the standard measure of semantic
relatedness in distributional semantics~\cite{TurneyPantel2010,Erk2012}. Spherical caps with radius $< \pi/2$
form good covers (Lemma~\ref{lem:good-cover-caps}), ensuring that the Čech nerve reflects the homotopy type
of the union via the classical Nerve Theorem~\cite{goerss2009simplicial}. The construction is thus fully
determined by established practice in both NLP and algebraic topology.

\medskip
\noindent\textbf{A runnable recipe (single slice).}
\begin{enumerate}\setlength\itemsep{0.3em}
  \item \emph{Embed and normalise.} Fix a frozen encoder $\Enc$ and a layer~$\ell$.
        Produce contextual embeddings $e_t$. Normalise to $\hat e_t \in S^{d-1}$.
        Obtain $P\subset S^{d-1}$ with metric $d_\angle$.
  \item \emph{Build caps.} Cluster under $d_\angle$. Pick centres $\mu_j$ and radii
        $\rho_j<\pi/2$. Record the incidence $R(t,j)$.
  \item \emph{Čech nerve.} $[j_0,\ldots,j_k]\in N(U)$ iff $\bigcap_r B_{j_r}\neq\varnothing$.
  \item \emph{(Optional) Filtration.} Sweep the radii to compute persistent homology.
        Prefer Čech; use VR as a proxy with caution.
  \item \emph{Fibrant replacement.} Set $A:=\ExInf N(U)$ for a compositional path calculus.
        Use $A$ for reasoning, $N(U)$ for measurement.
\end{enumerate}

\medskip
\noindent\textbf{Worked micro–example (no type theory yet).}
Let caps $B_{\mathsf{fin}}$, $B_{\mathsf{geo}}$, $B_{\mathsf{flow}}$ summarise ``bank'' (finance),
``bank'' (geomorphology), and ``flow''. Suppose $[\,\mathsf{fin},\mathsf{flow}\,]$ and
$[\,\mathsf{geo},\mathsf{flow}\,]$ are Čech edges but no triple intersection occurs at the sampling
radius. The measured $N(U)$ then has a $1$–cycle (triangle gap). In $A=\ExInf N(U)$, dashed short
composites witness connectivity for reasoning. The measured loop persists in $N(U)$ until a
larger radius introduces a $2$–simplex. Instantiating $\Enc$ with a fixed LLM, this toy example
corresponds to a situation where ``bank'' in river and finance contexts both remain linked through
flow-related usage. But the model has not yet produced a context that explicitly unifies all three.

%---------------------------------------------------------------
\section{Dynamic Homotopy Type Theory (DHoTT) calculus}
\label{sec:syntax-core}

\paragraph{Evolving text as data and as base.}
We fix a discrete time index set \(\Time\) (viewed as a small category with arrows
\(\tau'\!\to\!\tau\) for \(\tau\le\tau'\)). An \emph{evolving text} is given at each
instant \(\tau\) by a finite sequence of strings
\[
  \Corpus(\tau) \;=\; [\,s_1,\dots,s_{n_\tau}\,],
\]
for example the prefix of an ongoing dialogue with a large language model at
turn~\(\tau\). Concretely, in the conversational AI setting that motivates this work (recall Section~\ref{sec:intro}),
each $s_i$ is either a user prompt or a model response, and $\Corpus(\tau)$ is the concatenation of all
prompts and responses up to and including turn $\tau$. This growing sequence—what we called an ``evolving text''
in the introduction—is the primary object of study. At each turn the model reads $\Corpus(\tau)$, generates
a response, and this response is appended to form $\Corpus(\tau+1)$. Section~\ref{sec:example} will make this
concrete with an explicit four-turn dialogue.

A frozen encoder \(\Enc\) (as in Section~\ref{sec:measured-single-text})
yields contextual embeddings. We summarise these by a chosen map
\[
  \Embed_\tau:\Corpus(\tau)\longrightarrow S^{d-1}\subset \mathbb{R}^d
\]
and the angular metric \(\angle(-,-)\) on \(S^{d-1}\). From these measurements
we form a good cover \(U_\tau=\{B_\mu\}_{\mu\in K_\tau}\) by spherical caps. We build the Čech
nerve \(N(U_\tau)\). We take Kan fibrant replacement exactly as in
Section~\ref{sec:measured-single-text}. The resulting \emph{Evolving Text base}
at time \(\tau\) is
\[
  \ET(\tau) \;:=\; \Ex^\infty N(U_\tau) \;\in\; \SSet_{\mathrm{Kan}}.
\]
Thus $\ET(\tau)$ packages the measured geometry of the model's representation at turn $\tau$ into a
Kan complex suitable for HoTT reasoning.

\textit{Assumption (restriction coherence).}
We assume the measurement pipeline fixes a simplicial presheaf. The pipeline reads each \(\Corpus(\tau)\) through a
fixed \(\Enc\) and forms caps and Čech nerves. This produces
\[
  \ET : \Time^{\mathrm{op}} \longrightarrow \mathbf{sSet}_{\mathrm{Kan}},
\]
whose value at each $\tau$ is the Kan fibrant replacement $\ET(\tau) := \mathrm{Ex}^\infty N(U_\tau)$
of the Čech nerve of the measured cover. The restriction maps (the maps that relate later slices to earlier ones)
\[
  r_{\tau,\tau'} : \ET(\tau') \longrightarrow \ET(\tau)
\quad\text{for}\quad \tau \le \tau'
\]
satisfy
$r_{\tau,\tau} = \mathrm{id}$ and
$r_{\tau,\tau''} = r_{\tau,\tau'} \circ r_{\tau',\tau''}$.
In concrete implementations one may obtain such maps, for example, from a coherent
nearest-centre alignment of caps across time slices. The logic below uses only the
presheaf structure itself. It does not depend on a particular construction.

\paragraph{The ambient presheaf.}
Let \(\Time\) be a small poset of conversational times. We work in the presheaf topos
\[
  \DynSem \;:=\; [\,\Time^{\op}, \SSet\,],
\]
and regard \(\ET\) as a particular object of \(\DynSem\). Intuitively, \(\ET(\tau)\)
represents the measured Čech–based geometry of a model's embedding space at time \(\tau\).
As a Kan object it permits HoTT-style reasoning. To speak across time we use the
restriction maps \(r_{\tau,\tau'} : \ET(\tau')\to \ET(\tau)\). These flow ``backwards
in time'' and formalise how later material (including later LLM outputs) is interpreted
against earlier basins and caps.

\paragraph{Slices and fibres (correcting a common pitfall).}
We record the standard slice equivalence and preservation facts used throughout:

\begin{lemma}[Slices and fibres]
For every \(\tau\in\Time\) there is an equivalence of toposes
\[
  [\,\Time^{\op},\SSet\,]/y(\tau)\ \simeq\ [\,(\Time/\tau)^{\op},\SSet\,].
\]
Moreover, evaluation \(\mathrm{ev}_\tau : [\Time^{\op},\SSet]\to \SSet\) preserves
finite limits. Under the objectwise Kan–Quillen structure, it preserves fibrations
and weak equivalences.
\end{lemma}

\noindent\emph{Use.}
All ``local'' reasoning is done fibrewise. To study single-time geometry, we evaluate at
\(\tau\) and apply the usual simplicial-set facts. Good covers imply that Čech nerves
reflect shape; \(\Ex^\infty\) preserves homotopy type. Global, cross-time constructions
will later appear as natural transformations in~\(\DynSem\).

\paragraph{Ambient semantics.}
All reasoning is \emph{anchored in a slice}. Our calculus consists of judgements
\(\Gamma \vdash_{\tau} J\) to be read as:
``in the Kan fibre \(\ET(\tau)\), under assumptions \(\Gamma\) valid in that fibre,
the judgement \(J\) holds.'' Identity types are computed as path spaces in
\(\ET(\tau)\). All horn fillers needed for compositional path calculus exist
because \(\ET(\tau)\) is Kan.

We will form cross-time constructs (Carry, Rupture, and a provenance ledger) that may
mention \(\tau_0,\tau_1,\dots\) as \emph{parameters}. But the judgement itself is always
checked at a single anchor \(\tau\).

\paragraph{Universes.}
We assume a cumulative, univalent hierarchy in each fibre:
\[
  \U_0(\tau)\ \text{(small types at time \(\tau\))},\qquad
  \U_0(\tau)\in \U_1(\tau)\in \U_2(\tau)\cdots .
\]
A formation judgement \(\Gamma\vdash_\tau A:\U_0(\tau)\) means ``\(A\) is a small type in the Kan
fibre at \(\tau\).'' A term judgement \(\Gamma\vdash_\tau a:A\) means ``\(a\) is an element of
\(A\) in that fibre.'' Equality \(\Gamma\vdash_\tau \Id_A(a,b):\U_0(\tau)\) is the path type
in the fibre. When we speak about \emph{points and paths of the base itself}, we simply choose
the base as a fibre-type: \(x:\ET(\tau)\) and \(\Id_{\ET(\tau)}(x,y)\).

\paragraph{Contexts.}
A context \(\Gamma\) at time \(\tau\) is a telescope of fibre-types:
\[
  \Gamma \;\equiv\; x_1:A_1,\;
  x_2:A_2(x_1),\;\ldots,\;
  x_n:A_n(x_1,\ldots,x_{n-1}),
\]
where each \(A_i:\U_0(\tau)\) may depend on earlier variables in the telescope. Semantically,
this is a chain of Kan fibrations over the base \(\ET(\tau)\).

\subsection{Judgement forms}
We use a fibre-indexed Martin--Löf palette for each time $\tau$. We read \(\vdash_\tau\) as
``provable at time $\tau$ in the fibre'':
\[
\begin{array}{ll}
\Gamma \;\text{ctx}_\tau & \text{(a well-formed context in the fibre }\ET(\tau)\text{)} \\[2pt]
\Gamma \vdash_{\tau} A \;\text{type} & \text{(a type in the Kan slice }\ET(\tau)\text{)} \\[2pt]
\Gamma \vdash_{\tau} t : A & \text{(a term of }A\text{ at time }\tau\text{)} \\[2pt]
\Gamma \vdash_{\tau} \Id_A(x,y) \;\text{type}
 & \text{(identity as a path space in }\ET(\tau)\text{).}
\end{array}
\]
Inside a slice we keep the ordinary HoTT reading. Paths compose, invert, and act
functorially on dependents. Every path $\rho:x\to y$ in the base induces a transport
map in any dependent family $C:\ET(\tau)\to\mathcal{U}$.

\paragraph{Transport rule (fibrewise).}
At the level of judgments, transport along a path in $\ET(\tau)$ is given by the
standard HoTT rule, specialised to the base $A := \ET(\tau)$:
\begin{mathpar}
\inferrule*[right=transport]
  { \Gamma \vdash_{\tau} x : A \\
    \Gamma \vdash_{\tau} y : A \\
    \Gamma \vdash_{\tau} p : \Id_A(x,y) \\
    \Gamma,\, z{:}A \vdash_{\tau} C(z) : \mathcal{U} \\
    \Gamma \vdash_{\tau} u : C(x) }
  { \Gamma \vdash_{\tau} \transport_C(p,u) : C(y) }
\end{mathpar}
We apply this rule only to those families $C$ we intentionally interpret as
dependent types in the model. These are exactly the predicates that are stable under
the homotopy notion of identity in $\ET(\tau)$.

\paragraph{How \(\Gamma\) interacts with Kan replacement (semantics).}
Fix $\tau$.  A dependent family $C: \ET(\tau)\to \mathcal U$ is interpreted in
$[*,\SSet]$ by a \emph{Kan fibration} $p:\,E\to \ET(\tau)$ whose fibre over
$x\in \ET(\tau)$ is (a representative of) $C(x)$.  Equivalently, we write
$E \simeq \sum_{x:\ET(\tau)} C(x)$ for the Grothendieck construction (the total
space). The projection $\pi_1:\sum_{x:\ET(\tau)} C(x)\to \ET(\tau)$ is a fibration.
Because $\ET(\tau)$ is Kan and $p$ is a Kan fibration, every 1--simplex $\rho:x\to y$
in $\ET(\tau)$ admits path--lifting along $p$. The induced transport
\[
  \transport_C(\rho): C(x)\;\xrightarrow{\ \simeq\ }\; C(y)
\]
is the usual HoTT transport in the fibre $\ET(\tau)$. It is well defined up to the canonical
higher coherences.  In particular, for any such $C$, inhabitance of $C(x)$ and a path
$\rho:x\leadsto y$ in $\ET(\tau)$ forces inhabitance of $C(y)$. Internal predicates
are homotopy-constant on connected components of the base.

A context $\Gamma$ at time $\tau$ is interpreted as a finite composite of such
fibrations. Closure of Kan fibrations under pullback yields the standard structural
rules (weakening, exchange, contraction). The functoriality and associativity of
path--lifting in $\SSet_{\mathrm{Kan}}$ imply that reindexing along composites
$\rho_2\cdot\rho_1$ agrees judgmentally with successive reindexings.

\subsection{A type palette in the fibre (at time \texorpdfstring{$\tau$}{tau})}
\label{subsec:type-palette}
Here is an exemplary palette of useful fibre-types $A\in\U_0(\tau)$ that make internal
geometric reasoning explicit.  Throughout, $K_\tau$ denotes a finite index set of
spherical caps from the measured good-cover at $\tau$. For each $\mu\in K_\tau$ we
write $B_\mu$ for the external membership predicate ``point lies in cap $\mu$''. This is viewed
as part of the measured structure on the underlying vertices of $\ET(\tau)$.
These $B_\mu$ arise from the embedding geometry. They are \emph{not}, by default,
assumed to be preserved along arbitrary paths in $\ET(\tau)$.

\paragraph{(1) Base and labels.}
\begin{align*}
  \Base_\tau &\;\coloneqq\; \ET(\tau) \in \U_0(\tau)
    &&\text{(base points; allows $\Id_{\Base_\tau}$ paths).}\\
  \Label &:\ \Base_\tau \to \U_0(\tau)
    &&\text{(discrete labels attached to points, e.g.\ token id or surface string).}\\
  \Token &\;\coloneqq\; \Sigma(x:\Base_\tau).\ \Label(x) \in \U_0(\tau)
    &&\text{(labelled base-points).}
\end{align*}
\emph{Transport:} given $\rho:x\to y$ in $\Base_\tau$, dependent transport along $\rho$
yields $\transport_{\Label}(\rho):\Label(x)\to\Label(y)$. Similar transport holds for any
family over $\Base_\tau$ that we interpret as a Kan fibration.  These are the families
we intend to be \emph{stable} along semantic paths. Labels, tags, and other semantic
decorations are propagated along carried paths in the usual HoTT sense.

\paragraph{(2) Cap intersections (structural types).}
We do not treat raw cap-membership as an internal dependent family. But we can still
package cap intersections as genuine fibre-types by referring to the external
predicates $B_\mu$.  For a finite index set $I\subseteq K_\tau$, we define
\[
  \Simplex(I) \;\coloneqq\;
    \Sigma\bigl(x:\Base_\tau\bigr).\ \bigwedge_{\mu\in I} (B_\mu(x)=\mathtt{true})
    \;\in\; \U_0(\tau).
\]
An inhabitant of $\Simplex(I)$ is a \emph{witness} that the $I$-fold cap intersection
is nonempty. That is, the corresponding simplex exists \emph{here}.  These types expose
the measured good cover to the internal logic. They provide a bridge between the external
geometry ($B_\mu$) and internal reasoning. However, we do not use HoTT transport to
propagate $B_\mu$-truth values along arbitrary paths in $\Base_\tau$.

\paragraph{(3) Adjacency and chains.}
Using the external $B_\mu$, we can internalise adjacency and chains:
\begin{align*}
  \Edge(x,y) &\;\coloneqq\; \Sigma(\mu:K_\tau).\ (B_\mu(x)=\mathtt{true})\times(B_\mu(y)=\mathtt{true})
    \;\in \U_0(\tau),\\
  \PathAttempt(x,y) &\;\coloneqq\; \text{finite lists }[x=x_0,\ldots,x_n=y]\text{ with per-edge }\Edge(x_i,x_{i+1}).
\end{align*}
These may be optionally annotated with slack or angle bounds.  These are the raw materials for
admissibility predicates $\Adm$ used by Carry and Rupture.  Semantically they are
interpreted against the measured Čech+$Ex^\infty$ $1$--skeleton. An inhabitant of
$\Edge(x,y)$ is a certificate that $(x,y)$ is an edge. A term of
$\PathAttempt(x,y)$ is a concrete combinatorial chain.  We do not rely on these types
being invariant along long paths. They serve to expose the measured adjacency
structure to the internal calculus.

\paragraph{(4) Faces and horns (partial simplices).}
To talk about \emph{partial} combinatorics in the fibre:
\begin{align*}
  \Face(I,J) &\;\coloneqq\; (I\subseteq J)\times \Simplex(J),\\
  \OpenHorn(n,i) &\;\coloneqq\; \Sigma(\partial:\text{$n$-boundary data})\ .\ \neg\exists\,\Delta^n\!\to\!\ET(\tau)
\end{align*}
where the horn is indexed by $n\ge1$ and missing face $i$.
In practice we index $\OpenHorn$ by a failed $p\in\PathAttempt$ and record the missing
face. This is the positive, proof-relevant ``rupture payload'' we use in
\S\ref{subsec:rupture}.

\paragraph{(5) Properties over tokens (semantic decoration).}
Any $P:\Token\to \U_0(\tau)$ (POS tag, speaker, register, topic, and so on) is a legitimate
fibre-type.  These are the families that we \emph{intend} to be path-stable. For a
carried path $\rho:x\to y$ in $\Base_\tau$, the transport map
$\transport_P(\rho):P(x)\to P(y)$ propagates the decoration from $x$ to $y$
(see \S\ref{subsec:carry}).  This is the main way in which the internal logic of
$\ET(\tau)$ expresses semantic coherence along trajectories.

\paragraph{(6) Persistence representatives (for higher-level reasoning).}
For $k\in\{0,1,2\}$, a \emph{bar representative} is any explicit geometric witness
stored as a small type:
\[
  \RepBar^k \;\coloneqq\; \left\{
    \begin{array}{l}
      \text{tree with vertex list + edge evidence (for $k{=}0$),}\\[0.2em]
      \text{cycle with vertex list + 1-face evidence (for $k{=}1$),}\\[0.2em]
      \text{shell with face list + closure evidence (for $k{=}2$).}
    \end{array}
  \right.
  \qquad \RepBar^k\in \U_0(\tau).
\]
A bar representative can itself be made into a dependent family over $\Base_\tau$.
For example, ``token $x$ lies on bar $\beta$.'' In that case, transport along paths in
$\Base_\tau$ expresses the homotopy-invariance of that barcode witness.

\paragraph{Closure, hygiene, and external vs.\ internal predicates.}
The palette sits entirely in the fibre: $\U_0(\tau)$ is closed under $\Pi,\Sigma,\Id$
and standard HITs (computed fibrewise).  \emph{Measurement} produces the external
data $K_\tau$ and the predicates $B_\mu$. It induces the edge relation. \emph{Reasoning}
manipulates the resulting internal fibre-types such as $\Base_\tau$, $\Token$,
$\RepBar^k$, and semantic decorations $P:\Token\to\U_0(\tau)$.  Transport is always
computed in the Kan fibre $\ET(\tau)$ along paths $\rho:\Id_{\ET(\tau)}(x,y)$. But we
reserve it for those families $C$ we explicitly interpret as Kan fibrations. That is,
for predicates we choose to regard as invariant under the notion of semantic
coherence encoded by paths in $\ET(\tau)$.

\subsection{Core Martin--Löf rules}

Standard HoTT constructs ($\Pi$, $\Sigma$, $\Id$) remain unchanged. For brevity, we recall only
the $\Pi$-formation rule explicitly:
\begin{mathpar}
\inferrule*[right=$\Pi$-Formation]
  { \Gamma \vdash_{\tau} A\;\text{type} \quad
    \Gamma, x{:}A \vdash_{\tau} B(x)\;\text{type} }
  { \Gamma \vdash_{\tau} \Pi_{x:A} B(x)\;\text{type} }
\end{mathpar}

\subsection{Carry (semantic continuity across time)}
\label{subsec:carry}

\paragraph{Reading.}
In this paper we define cross-time carry at the level of the \emph{base} $\ET(\tau)$ and then
allow dependent families over $\ET(\tau)$ to come along for the ride. A \emph{carry witness}
from $\tau$ to $\tau'$ does not assert an in-slice equality; it certifies that a base point
$a:\ET(\tau)$ reappears coherently at $\tau'$ by providing a later echo $a' : \ET(\tau')$
together with a path \emph{in the earlier slice} from $a$ to the earlier view of $a'$.
Internally we use the restriction map $r_{\tau,\tau'} : \ET(\tau') \to \ET(\tau)$ and write
paths in $\ET(\tau)$ using $\Id_{\ET(\tau)}(\,\cdot\,,\,\cdot\,)$.

\paragraph{Formation (family of carry certificates).}
Given a base point $a:\ET(\tau)$, we form the type of carry certificates from $\tau$ to
$\tau'$ for $a$:
\[
  \Carry_{\ET}^{\tau\to\tau'}(a)
  \;:\; \text{Type}
  \;\overset{\mathrm{def}}{:=}\;
  \Sigma\bigl(a' : \ET(\tau')\bigr)\; \Id_{\ET(\tau)}\!\bigl(r_{\tau,\tau'}(a'),\,a\bigr).
\]
Judgmentally:
\begin{mathpar}
\inferrule*[right=Carry-Form]
  { \Gamma \vdash_{\tau} a : \ET(\tau) \quad \tau \le \tau' }
  { \Gamma \vdash_{\tau} \Carry_{\ET}^{\tau\to\tau'}(a)\;\text{type} }
\end{mathpar}

\paragraph{Introduction (exhibiting a carry).}
A carry is introduced by exhibiting a later echo $a'$ and a path $\rho$ in the \emph{earlier}
slice:
\begin{mathpar}
\inferrule*[right=Carry-Intro]
  { \Gamma \vdash_{\tau'} a' : \ET(\tau') \quad
    \Gamma \vdash_{\tau} \rho : \Id_{\ET(\tau)}\!\bigl(r_{\tau,\tau'}(a'),\,a\bigr) }
  { \Gamma \vdash_{\tau} \langle a',\rho\rangle : \Carry_{\ET}^{\tau\to\tau'}(a) }
\end{mathpar}

\paragraph{Elimination / use (transport along a carry).}
Let $C : \ET(\tau) \to \U_0(\tau)$ be a dependent family (at time $\tau$). Define
$C^{\uparrow[\tau\to\tau']} : \ET(\tau') \to \U_0(\tau)$ by
\[
  C^{\uparrow[\tau\to\tau']}(a') := C\!\bigl(r_{\tau,\tau'}(a')\bigr).
\]
Then a carry certificate allows dependent transport:
\begin{mathpar}
\inferrule*[right=Carry-Use]
  { \Gamma \vdash_{\tau} a : \ET(\tau) \quad
    \Gamma \vdash_{\tau} t : C(a) \quad
    \Gamma \vdash_{\tau} \langle a',\rho\rangle : \Carry_{\ET}^{\tau\to\tau'}(a) }
  { \Gamma \vdash_{\tau'} \ 
      \underbrace{\vphantom{\Big|}\smash{\,\transport_{C}(\rho)(t)\,}}_{\text{``carry-use''}}
      \;:\; C^{\uparrow[\tau\to\tau']}(a') }
\end{mathpar}
By definition, $\transport_{C}(\rho)(t)$ is the ordinary HoTT transport along $\rho$ in the Kan
fibre $\ET(\tau)$; we simply \emph{read} the result as attached to the later echo $a'$ via
$C^{\uparrow[\tau\to\tau']}(a') = C(r_{\tau,\tau'}(a'))$.

\paragraph{Computation (collapse to HoTT transport).}
When $\tau=\tau'$ we have $r_{\tau,\tau}=\mathsf{id}$, so
\[
  \Carry_{\ET}^{\tau\to\tau}(a)\;\simeq\;\Sigma\bigl(a':\ET(\tau)\bigr)\,\Id_{\ET(\tau)}(a',a).
\]
In particular, for the \emph{trivial} carry $\langle a,\refl_a\rangle$ and any $t{:}C(a)$,
\[
  \transport_{C}(\refl_a)(t)\; \equiv\; t.
\]
Thus carry is strictly stronger data than in-slice transport, but \emph{collapses} to it on
fixed slices.

\paragraph{Composition of carries.}
Carries compose via functoriality of restriction and path composition in the earlier slice.
\begin{mathpar}
\inferrule*[right=Carry-Comp]
  { \kappa_{01}=\langle a_1,\rho_{01}\rangle : \ \Carry_{\ET}^{\tau_0\to\tau_1}(a_0)
    \quad
    \kappa_{12}=\langle a_2,\rho_{12}\rangle : \ \Carry_{\ET}^{\tau_1\to\tau_2}(a_1) }
  { \Bigl\langle a_2,\ \rho_{01}\cdot r_{\tau_0,\tau_1}(\rho_{12}) \Bigr\rangle
      : \ \Carry_{\ET}^{\tau_0\to\tau_2}(a_0) }
\end{mathpar}
Here $r_{\tau_0,\tau_1}(\rho_{12})$ is the image of the later path under restriction
$\ET(\tau_1)\to \ET(\tau_0)$, and $\cdot$ denotes path concatenation in
$\Id_{\ET(\tau_0)}$. Moreover, for any $t\!:\!C(a_0)$ one has the judgmental equality
\[
  \transport_{C}\!\bigl(\rho_{01}\!\cdot\! r_{\tau_0,\tau_1}(\rho_{12})\bigr)(t)
  \;\equiv\;
  \transport_{C^{\uparrow[\tau_0\to\tau_1]}}\!\bigl(\rho_{12}\bigr)\!\Bigl(\, \transport_{C}(\rho_{01})(t)\,\Bigr),
\]
i.e.\ transport along a composite carry equals sequential transport.

\paragraph{Units.}
For every $\tau$ and $a:\ET(\tau)$ there is a unit carry
$\mathsf{ucarry}(a)\;:=\;\langle a,\refl_a\rangle : \ \Carry_{\ET}^{\tau\to\tau}(a)$.
These are left/right units for the composition above, up to the standard groupoid coherences
in $\ET(\tau)$.

\subsection{Dependent carry}
\label{subsec:dep-carry}

Families reindex along their base via restriction and ordinary transport in the earlier slice.

\paragraph{Formation (families come along for the ride).}
If $\Gamma \vdash_{\tau} a:\ET(\tau)$ and $\Gamma, x{:}\ET(\tau) \vdash_{\tau} P(x)\;\text{type}$,
a family over $\ET(\tau)$ at time $\tau$ induces a family over carry certificates by
\[
  \Carry_{P}^{\tau\to\tau'}\!\bigl(\langle a',\rho\rangle\bigr)
  \;:\; \text{Type}
  \;\overset{\mathrm{def}}{:=}\;
  P\!\bigl(r_{\tau,\tau'}(a')\bigr).
\]

\paragraph{Transport (dependent terms).}
Given $\Gamma, x{:}\ET(\tau) \vdash_{\tau} t(x):P(x)$ and
$\kappa= \langle a',\rho\rangle : \ \Carry_{\ET}^{\tau\to\tau'}(a)$, we obtain:
\begin{mathpar}
\inferrule*[right=Dep-Carry-Transp]
  { \Gamma \vdash_{\tau} a : \ET(\tau) \quad
    \Gamma, x{:}\ET(\tau) \vdash_{\tau} t(x):P(x) \quad
    \Gamma \vdash_{\tau} \langle a',\rho\rangle : \ \Carry_{\ET}^{\tau\to\tau'}(a) }
  { \Gamma \vdash_{\tau'} \ \transport_{P}(\rho)\bigl(t(a)\bigr) : P\!\bigl(r_{\tau,\tau'}(a')\bigr) }
\end{mathpar}
Dependent carry respects identities and composition as above; if $P$ is constant then
$\transport_{P}(\rho)$ is judgmentally the identity.

\subsection{Rupture}
\label{subsec:rupture}

A cross–time move may \emph{fail} to admit, under a fixed fibrewise admissibility policy
\(\Adm\), an admissible carry for \(a\in \ET(\tau)\) towards a later slice \(\tau'\).
In our calculus, such failure is not a mere negation but a witnessed proof object that
records what was tried and where it broke. We call this the \emph{rupture}.

\paragraph{Policy and attempts (in the earlier fibre).}
We fix \(\tau\le \tau'\). All checks occur in the \emph{earlier} fibre \(\ET(\tau)\).
Let \(\PathAttempt(a,b)\) denote finite Čech–guided chains in \(\ET(\tau)\) that try to
connect \(a\) to \(b\). These chains carry per-edge evidence (for example, shared-cap witnesses, slack bounds, or angle bounds). Let \(\OpenHorn(p)\) tag the unfilled face or horn identified by the attempt \(p\).
An \emph{admissibility policy} \(\Adm\) is a decidable predicate on attempts. We assume it is \emph{local}, \emph{finite},
\emph{monotone} under cover refinement, and stable under restriction maps \(r_{\tau,\tau'}\).

\begin{example}[Concrete policy]\label{ex:concrete-policy}
A typical policy might require: $\Adm([j_0,\ldots,j_n]) = \mathtt{true}$ if and only if
(1) the chain length satisfies $n \le 3$, and
(2) each per-edge angle satisfies $\angle(\mu_{j_i}, \mu_{j_{i+1}}) < \pi/6$.
This policy admits only short chains with small angular steps.
\end{example}

\paragraph{Finite attempt space (enumerable under policy).}
At a fixed anchor time $\tau$, the cover $U_\tau=\{B_\mu\}_{\mu\in K_\tau}$ is finite, and
the admissibility policy $\Adm$ uses only local, decidable checks (bounded hop $H$,
overlap/angle thresholds, \dots). We define attempts over a finite search graph in the
\emph{earlier} fibre and cap search at $H$.

\medskip\noindent\emph{Cap–index variant.}
Let $G_\tau^{\mathrm{cap}}=(K_\tau,E)$ with $(\mu,\nu)\in E$ iff $B_\mu\cap B_\nu\neq\varnothing$
and the overlap meets the policy thresholds. For $a\in\ET(\tau)$ and $a'\in\ET(\tau')$,
\[
  \PathAttempt(a, r_{\tau,\tau'}(a'))
  \;:=\;
  \Bigl\{\, [\mu_0,\ldots,\mu_n]\ \Big|\ 
  0\le n\le H,\ a\in B_{\mu_0},\ r_{\tau,\tau'}(a')\in B_{\mu_n},\ 
  (\mu_i,\mu_{i+1})\in E\ \forall i
  \Bigr\}.
\]
The policy $\Adm([\mu_0,\ldots,\mu_n])$ is a decidable conjunction of local constraints
(e.g.\ $n\!\le\!H$, per–edge angle $\le\delta_{\mathrm{eff}}$, required cap overlaps).

\medskip\noindent\emph{Witness–point variant.}
Alternatively, let $G_\tau^{\mathrm{pt}}=(V,E)$ with $V$ a finite witness set of base
points in $\ET(\tau)$ (e.g.\ measured occurrences or chosen representatives), and
$(x,y)\in E$ when $\exists \mu\in K_\tau$ with $x,y\in B_\mu$ and
$\angle(\Embed_\tau(x),\Embed_\tau(y))\le\delta_{\mathrm{eff}}$. Define
$\PathAttempt(a, r_{\tau,\tau'}(a'))$ as bounded walks $[x_0,\ldots,x_n]$ with
$x_0$ witnessing $a$ and $x_n$ witnessing $r_{\tau,\tau'}(a')$.

\medskip
Both variants are inter-definable in the finite setting; choose whichever is
convenient for proofs or implementation. In an LLM setting, $\Adm$ might encode
a policy such as ``paths no longer than $H$ and angles below a drift threshold
$\delta_{\mathrm{eff}}$'' when diagnosing semantic drift or hallucinated jumps.

\begin{lemma}[Finiteness and decidability of attempts]
With finite $U_\tau$, hop bound $H$, and local decidable $\Adm$, the set
$\PathAttempt(a, r_{\tau,\tau'}(a'))$ is finite and effectively enumerable
(e.g.\ by BFS/DFS with pruning). Consequently, the rupture payload
\[
  \Rupture_{\ET}(\tau\!\to\!\tau';a)
  \;=\;
  \sum_{a' : \ET(\tau')}\ \sum_{p\in\PathAttempt(a, r_{\tau,\tau'}(a'))}
  \bigl(\neg\Adm(p)\bigr)\times \OpenHorn(p)
\]
is a small fibre–type in $\U_0(\tau)$, and appending entries to
$\RupLed_{\ET}^{\Adm}(\tau)(a)$ is decidable.
\end{lemma}

In the cap-index variant, we explore the finite graph
$G^{\mathrm{cap}}_\tau = (K_\tau, E)$ by breadth-first search truncated at hop
bound $H$.  The resulting search space is finite and effectively enumerable, with
size determined by $|K_\tau|$, $|E|$, and $H$; we do not rely on any sharper
asymptotic complexity bound.  The witness-point variant enjoys an analogous
property with $|V|$ in place of $|K_\tau|$.

\paragraph{Internal proof objects for rupture.}
At the measurement level, failure at $\tau$ is decidable for the geometric,
embedding-based semantics used to build the Čech nerve (cf.\
Section~\ref{sec:measured-single-text}): given a finite graph and a fixed
admissibility policy $\Adm$, we can simply search for bounded-hop paths and
see whether any succeed.  So we do not \emph{need} a new logical constructor
in order to tell whether a particular attempt has failed.  However, in software
engineering as in constructive logic, logs and audit trails are crucial for
understanding and explaining system behaviour.  The same applies in DHoTT as a
language for evolving textual meaning.  Internalising rupture as a fibre-type
(i.e.\ as a proof-relevant $\Sigma$-object) yields:
(i) \emph{provenance} as first-class terms that can be transported, compared, and
referenced in further derivations;
(ii) a \emph{stable interface} if $\Adm$ later takes another form (for example,
human-in-the-loop audits or ML retraining that changes coherence paths by
re-weighting the embedding model and thereby introducing new admissible edges);
and
(iii) \emph{auditability} across policy/version changes without altering the
proof rules, since the ledger stores the original failure witnesses rather than
only their Boolean shadow.

\paragraph{Canonical rupture (payloaded).}
The rupture type between \(\tau\) and \(\tau'\) for \(a\in\ET(\tau)\) is the small fibre-type
\[
  \Rupture_{\ET}(\tau\!\to\!\tau';a)
  \;\;:=\;\;
  \sum_{a' : \ET(\tau')}\;\sum_{p:\PathAttempt\!\bigl(a,\;r_{\tau,\tau'}(a')\bigr)}
  \Bigl(\neg\,\Adm(p)\Bigr)\times \OpenHorn(p)
  \;\in\; \U_0(\tau).
\]
An inhabitant is a \emph{structured failure}:
a later candidate \(a'\), a concrete attempt \(p\) (anchored in \(\ET(\tau)\)), and a
witness of why it fails (policy violation + open-horn tag). Elimination is by the
ordinary $\Sigma$-eliminator; we do not need a specialised rule.

\paragraph{Carry and use (recall).}
A \emph{carry} is (as in \S\ref{subsec:carry})
\[
  \Carry_{\ET}^{\tau\to\tau''}(a)
  \;:=\;
  \sum_{a'':\ET(\tau'')}\ \Id_{\ET(\tau)}\!\bigl(r_{\tau,\tau''}(a''),\,a\bigr),
\]
and carry-use is ordinary HoTT transport along the path component in \(\ET(\tau)\).
Rupture values never fabricate a carry; operations that require a carry remain inapplicable
until a \(\kappa\in\Carry_{\ET}^{\tau\to\tau''}(a)\) is actually present.

\paragraph{Rupture ledger (provenance, append-only).}
To record the history of failures (and eventual success), we keep an append-only ledger per
anchor \(a\):
\[
  \RupLed_{\ET}^{\Adm}(\tau_0)(a)
  \;:=\;
  \mathsf{List}\!\left(
    \sum_{\tau'\ge\tau_0}\ \Rupture_{\ET}(\tau_0\!\to\!\tau';a)
    \;\;+\;\;
    \sum_{\tau''\ge\tau_0}\ \Carry_{\ET}^{\tau_0\to\tau''}(a)
  \right)
  \;\in\; \U_0(\tau_0).
\]
\emph{Constructors:} \(\mathsf{empty}\) and \(\mathsf{append}\).
Entries are time-stamped and only appended; past rupture payloads are never erased.
When a later carry \(\kappa=\langle a'',\rho\rangle\) is found, we append $\mathsf{success}(\kappa)$.
(If desired for exposition, one may define a derived macro \(\ExposeGlue(\kappa)\) that names
the canonical gluing 1-simplex associated to \(\rho\); this is presentation sugar, not a rule.)

\paragraph{Honesty and blocking.}
Let \(\NoCarryYet_{\ET}^{\Adm}(\tau\!\to\!\tau';a)\) be the propositional statement
\(\neg\exists \kappa\in \Carry_{\ET}^{\tau\to\tau'}(a)\) whose witness satisfies $\Adm$.
This proposition \emph{blocks} rules that require a carry; rupture payloads
\(\Rupture_{\ET}\) provide constructive \emph{evidence of attempts}, but do not by themselves
entail blocking or unblocking. In practice, search produces rupture entries; when it finally
produces \(\kappa\), carry-use proceeds by $\transport$ in \(\ET(\tau)\).

\paragraph{Semantics.}
\(\Rupture_{\ET}(\tau\!\to\!\tau';a)\) is interpreted \emph{objectwise} in
\(\SSet_{\mathrm{Kan}}\) as the displayed \(\Sigma\)–type: later candidate \(a'\),
attempted chain \(p\) in \(\ET(\tau)\), policy failure, and an open-horn tag in the measured
skeleton. There is no homotopy-pushout carrier here; the only gluing we ever mention is the
\emph{derived} canonical 1-simplex that accompanies a \emph{later} carry's path component,
used purely for diagrammatic exposition when desired. All metatheory (substitution, soundness,
fibrancy) is fibrewise and follows from closure of \(\U_0(\tau)\) under \(\Sigma,\Pi,\Id\).

\paragraph{Stability and decidability of policy.}
If \(\Adm\) depends only on finitely many local checks (bounded chain length, overlap/angle
thresholds) and is monotone under refinement, then for fixed \(a,\tau\) the predicate
\(\NoCarryYet_{\ET}^{\Adm}(\tau\!\to\!(-); a)\) is monotone in \(\tau'\) and \emph{decidable}.
Hence ledger appends are well-defined; introducing a later success is a conservative update of
data, not a change of rules.

\begin{remark}[Measurement vs.\ reasoning]
\emph{Measurement} precedes logic:
\[
  \text{strings}\ \to\ \ell_2\text{-embeddings on }S^{d-1}\ \to\ \text{good cover }U_\tau\ \to\ N(U_\tau)\ \to\ \Ex^\infty N(U_\tau).
\]
\emph{Reasoning} occurs \emph{in} the fibre \(\ET(\tau)\): identities are paths; horn fillers
license composition; and dependent transport is the ordinary HoTT transport inside the fibre.
This hygiene ensures we never ``measure our inferences.''
\end{remark}

\begin{remark}[No adjacency; patience is internal]
Nothing in the rules assumes that $\tau'$ is a ``next step.'' The time index $\Time$ can
be any discrete sampling of an evolving text or LLM-driven conversation: successive
turns, windows, or even hand-picked checkpoints. The carry and rupture rules are
agnostic to how densely we sample.

Semantically, over an evolving text, two things may happen to a token's meaning as we
move from $\tau$ to later times:
\begin{itemize}
  \item \emph{Disappearance.} The token (or its surface form) simply no longer occurs in
        $C(\tau')$ for some $\tau'>\tau$. In that case there is nothing to carry, and no
        rupture is formed; the ledger for $a\in\ET(\tau)$ remains unchanged.
  \item \emph{Rupture.} The token does reappear in $C(\tau')$, but every admissible
        attempt $p\in\PathAttempt(a,r_{\tau,\tau'}(a'))$ fails the policy $\Adm$. In that
        case we record a structured failure in $\Rupture_{\ET}(\tau\!\to\!\tau';a)$ and
        append it to the ledger $\RupLed_{\ET}^{\Adm}(\tau)(a)$.
\end{itemize}
Carries compose (cf.\ \textsc{Carry-Comp}) and ledger-based ``healing'' via \emph{later}
carries is conservative regardless of how large $\tau''$ is: if at some much later time
$\tau''\gg\tau$ we finally obtain $\kappa=\langle a'',\rho\rangle\in
\Carry_{\ET}^{\tau\to\tau''}(a)$, then the ledger simply acquires a $\mathsf{success}(\kappa)$
entry, and all transport remains computed in the original fibre $\ET(\tau)$.

From the point of view of distributional semantics, this yields a constructive account of
``meaning as use'' over time. The evolving geometry of contextual embeddings produced by a
fixed encoder $\Enc$ is not read off from raw distances alone, but from \emph{witnessed}
paths and witnessed failures: carries certify that a later use of a token is still reachable
from an earlier one under a chosen policy $\Adm$; ruptures certify that, under the same
policy, all finite attempts fail. The append-only ledger then records when a sign's meaning
drifts out of reach and when (if ever) it \emph{re-enters} via a later carry.

In this sense DHoTT provides a logical kernel for the geometry of meaning evolution in the
distributional sense. It packages empirical observations about LLM hidden states or other
embedding spaces (via Čech covers and nerves) into proof-relevant objects (carries,
ruptures, ledgers) inside a type theory. This opens a path toward formalising currently
purely experimental approaches to analysing coherence and hallucination in large language
models, which are all ultimately based on observing how embeddings move in time.
\end{remark}

\subsection{Metatheoretic properties}
\label{subsec:metatheory-revised}

\begin{theorem}[Syntactic substitution]\label{theorem:subst-syntax}
If $\Gamma\vdash_{\tau} J$ and $\sigma:\Delta\to\Gamma$, then $\Delta\vdash_{\tau} J[\sigma]$.
\end{theorem}

\begin{proof}
By induction on the derivation of $\Gamma\vdash_{\tau}J$.
Structural cases use pullback naturality.
Core HoTT cases are standard. For the new constructors:

\emph{Carry-Form/Use}: $\Carry_{\ET}^{\tau\to\tau'}(a)\;=\;
\Sigma(a':\ET(\tau'))\,.\,\Id_{\ET(\tau)}(r_{\tau,\tau'}(a'),a)$
is built from $\Sigma$ and $\Id$ in the fibre $\ET(\tau)$, so it commutes strictly with
substitution. Carry-Use evaluates to ordinary transport along the path $\rho$ in $\ET(\tau)$,
hence is strictly natural.

\emph{Rupture-Form}: $\Rupture_{\ET}(\tau\!\to\!\tau';a)$ is a finite $\Sigma$-type built
from $\ET(\tau')$, $\PathAttempt$, $\neg\Adm$, and $\OpenHorn(p)$ in the fibre $\ET(\tau)$,
so it also commutes strictly with substitution. Elimination is the ordinary $\Sigma$-eliminator.
\end{proof}

\begin{remark}[Partial canonicity]\label{rmk:partial-canonicity}
Any closed term of a base inductive type that does \emph{not} involve rupture
coherences reduces to a canonical constructor. Full canonicity for terms that
\emph{do} mention rupture paths is left open (as in standard HoTT with general HITs).
\end{remark}

% ============================================================
\section{Semantics}\label{sec:semantics}

Our canonical model is the simplicial–presheaf topos
\[
  \DynSem \;:=\; [\,\Time^{\op},\,\SSet\,],
\]
where \(\Time\) is a small poset of times and we use the projective (objectwise Kan–Quillen)
model structure. Limits and colimits are computed pointwise. For each \(\tau\in\Time\),
evaluation at \(\tau\)
\[
  \ev_\tau : \DynSem \longrightarrow \SSet
\]
is right adjoint to base–change along \(y(\tau)\), hence preserves finite limits,
fibrations, and weak equivalences. Slices behave in the standard way:
\[
  [\,\Time^{\op},\SSet\,]/y(\tau)\ \simeq\ [\,(\Time/\tau)^{\op},\SSet\,].
\]
Consequently, reasoning over a single time \(\tau\) is carried out \emph{fibrewise}
in the Kan slice \(\SSet_{\mathrm{Kan}}\) via \(\ev_\tau\). Univalence and the
standard HoTT constructors hold \emph{objectwise}; we never identify the slice topos
with \(\SSet\) itself.

\subsection{Interpretation of judgements}\label{subsec:interp-revised}

Fix a Grothendieck universe bound so that all simplicial sets we form are small.
Interpretation is by induction on derivations:
\[
  \Gamma \mapsto \llbracket \Gamma \rrbracket \in \DynSem,\qquad
  \Gamma\vdash A \mapsto \llbracket A \rrbracket :
  \llbracket \Gamma \rrbracket \to \mathcal{U},
\]
with contexts as telescopes (dependent products computed objectwise in \(\SSet\)).

\paragraph{Core type formers.}
The core type formers $\Pi,\Sigma,\Id$ and the standard higher inductive types are
interpreted \emph{objectwise} in $\mathbf{sSet}$.  Local (single-time) reasoning is
therefore performed by evaluating at $\tau$ via the evaluation functor
$\mathrm{ev}_\tau \colon [\Time^{\mathrm{op}},\mathbf{sSet}] \to \mathbf{sSet}$ and working
in the Kan complex $ET(\tau)$.  For universes we simply reuse any standard simplicial
univalent universe of Kan complexes; concretely, we may take a univalent universe
$\mathcal{U}\in\mathbf{sSet}$ from the Kapulkin–Lumsdaine simplicial model of
univalent foundations \cite{HoTTBook}, and interpret our universe presheaf
by setting $\mathcal{U}(\tau) := \mathcal{U}$ for each $\tau$.  Thus each fibre
$\mathcal{U}(\tau)$ is a univalent universe in the usual sense, and univalence holds
fibrewise across $\mathsf{DynSem}$; we do not require or assume a separate global
univalent universe in the ambient presheaf topos.

\paragraph{Carry.}
For the base object $\ET:\Time^{\op}\to\SSet$, the semantic interpretation of
$\Carry_{\ET}^{\tau\to\tau'}(a)$ for $a\in\ET(\tau)$ is exactly the $\Sigma$-type used in the
rules:
\[
  \big\llbracket\Carry_{\ET}^{\tau\to\tau'}(a)\big\rrbracket
  \;:=\;
  \Sigma\bigl(a':\ET(\tau')\bigr)\,.\,\Id_{\ET(\tau)}\!\bigl(r_{\tau,\tau'}(a'),\,a\bigr),
\]
i.e.\ a later echo \(a'\) together with a path \emph{in the earlier fibre} \(\ET(\tau)\).
Carry–Use is ordinary HoTT transport along that path inside \(\ET(\tau)\); composition
agrees with sequential transport via restriction \(r_{\tau,\tau'}\) and path concatenation.

\paragraph{Rupture (constructive semantics, no carriers).}
We interpret rupture as the \emph{positive} $\Sigma$–type used in the rules (no pushouts).
Fix \(\tau\le\tau'\), the base \(\ET:\Time^{\op}\to\SSet\), an anchor \(a\in\ET(\tau)\),
and a fixed local, finite, decidable policy \(\Adm\) in the earlier fibre \(\ET(\tau)\).
Let \(\PathAttempt(a,b)\) denote the \emph{finite} set of bounded Čech–chains in \(\ET(\tau)\)
that try to connect \(a\) to \(b\), and \(\OpenHorn(p)\) the tag of the missing face/horn
determined by \(p\). Then, \emph{objectwise in the fibre at \(\tau\)}:
\begin{multline*}
      \big\llbracket \Rupture_{\ET}(\tau\!\to\!\tau';a) \big\rrbracket
  \;:=\;
  \Sigma\Bigl(a' : \ET(\tau')\Bigr)\ \Sigma\Bigl(p:\PathAttempt\!\bigl(a, r_{\tau,\tau'}(a')\bigr)\Bigr) 
  \\ \Bigl(\neg\,\Adm(p)\Bigr)\times \OpenHorn(p).
\end{multline*}
Thus a rupture is an \emph{internal provenance object}: a later candidate,
a concrete attempt \(p\) anchored in \(\ET(\tau)\), and a proof–relevant reason
it fails. There is no homotopy–pushout ``carrier''; the only gluing we ever mention
is the \emph{derived} canonical 1–simplex associated to a \emph{later} carry's path
component, used purely for diagrammatic exposition when desired.

\paragraph{Attempt space and smallness (objectwise).}
At a fixed \(\tau\), the cover \(U_\tau\) is finite and we fix a hop bound \(H\).
We build a finite search graph in \(\ET(\tau)\) (cap–index or witness–point variant)
and define \(\PathAttempt\) as bounded walks; \(\Adm\) is a decidable conjunction
of local constraints (length \(\le H\), overlap/angle thresholds, \dots).
Hence \(\PathAttempt\) is finite and enumerable objectwise, so
\(\Rupture_{\ET}(\tau\!\to\!\tau';a)\) is a small fibre–type in \(\U_0(\tau)\).
Ledger updates (\(\RupLed\)) are likewise decidable and interpreted objectwise
as lists in \(\SSet\).

\paragraph{Soundness, fibrancy, substitution.}
All clauses above are built from objectwise \(\Sigma,\Pi,\Id\) and finite families;
\(\Carry\) and \(\Rupture\) preserve Kan fibrancy fibrewise, and substitution is
interpreted by pullback/naturality of evaluation \(\ev_\tau\).
No additional HITs are required by rupture; all metatheory reduces to standard
simplicial–set arguments in each fibre.

\subsection{Fibrancy and soundness}\label{subsec:fibrancy-soundness}

\begin{lemma}[Fibrancy]\label{lem:fibrancy-revised}
For every derivable judgement $\Gamma\vdash_\tau A:\text{type}$, the map
$\llbracket A \rrbracket\to\llbracket\Gamma\rrbracket$ is a small fibration in the
projective model structure on $\DynSem=[\Time^{\op},\SSet]$.
\end{lemma}

\begin{proof}
By induction on formation. Core HoTT cases are interpreted objectwise in $\SSet$,
hence fibrant. For \emph{Carry}, the interpretation is
$\Sigma(a':\ET(\tau')).\,\Id_{\ET(\tau)}(r_{\tau,\tau'}(a'),a)$, so $\Sigma$ and $\Id$
preserve fibrancy objectwise. For \emph{Rupture}, the interpretation is a \emph{finite}
objectwise $\Sigma$ over $\ET(\tau')$ and the finite set $\PathAttempt$ in the earlier
fibre, followed by a decidable predicate and a tag $\OpenHorn(p)$:
\[
  \big\llbracket \Rupture_{\ET}(\tau\!\to\!\tau';a) \big\rrbracket
  \;=\;
  \Sigma\Bigl(a' : \ET(\tau')\Bigr)\ \Sigma\Bigl(p:\PathAttempt(a,r_{\tau,\tau'}(a'))\Bigr)
  \ \Bigl(\neg\,\Adm(p)\Bigr)\times \OpenHorn(p).
\]
Small $\Sigma$–types and finite products preserve fibrancy objectwise in $\SSet$.
Hence the presheaf is fibrant.
\end{proof}

\begin{theorem}[Soundness]\label{theorem:soundness-revised}
Every derivable judgement $\Gamma\mid\tau\vdash J$ is interpreted by a well-typed
morphism in $\DynSem$ that satisfies the corresponding computation rules.
\end{theorem}

\begin{proof}
Structural induction on derivations. Core HoTT: standard for simplicial presheaves.
\emph{Carry-Form/Use}: immediate from the $\Sigma/\Id$ semantics and ordinary HoTT transport
in the earlier fibre. \emph{Rupture}: by the universal property of $\Sigma$ and the
objectwise interpretation of $\PathAttempt$, $\neg\Adm$, and $\OpenHorn(p)$ in $\SSet$.
\end{proof}

\begin{corollary}[Semantic substitution]\label{cor:subst-revised}
For any context morphism $\sigma:\Delta\to\Gamma$ and judgement
$\Gamma\mid\tau\vdash J$, we have
\(
\llbracket J[\sigma]\rrbracket
=
\llbracket J\rrbracket\circ\llbracket\sigma\rrbracket.
\)
\end{corollary}

\begin{proof}
Simultaneous induction on $\sigma$ and $J$. All interpreting operations
(pullback, $\Pi,\Sigma,\Id$, and the objectwise constructions used by \emph{Carry} and \emph{Rupture})
commute strictly with base-change in the projective model structure. Carry-Use reduces to
naturality of HoTT transport in the fibre $\ET(\tau)$.
\end{proof}

\begin{theorem}[Conservativity]\label{theorem:conservativity-revised}
If $J$ is a closed HoTT judgement, then $\mathrm{HoTT}\vdash J$ iff $\mathrm{DHoTT}\vdash J$.
\end{theorem}

\begin{proof}
Embed HoTT at a fixed $\tau$ (left-to-right), and evaluate any DHoTT derivation at a fixed $\tau$
(right-to-left); evaluation preserves the univalent universe and all structure objectwise in $\SSet$.
\end{proof}

\begin{proposition}[Fibrewise univalence and naturality]\label{prop:fibrewise-ua}
For each $\tau$, the universe $\mathcal{U}(\tau)$ is univalent (simplicial sets);
restriction $r_{\tau,\tau'}$ preserves equivalences objectwise. Thus equivalences
are stable under time restriction; no extra ``temporal univalence'' principle is needed.
\end{proposition}

\begin{lemma}[Carry–substitution commutes strictly]\label{lem:carry-subst}
Let $\Gamma,x{:}\ET(\tau)\vdash_{\tau} t:C$ and $\Gamma\vdash_{\tau} a:\ET(\tau)$.
If $\kappa=\langle a',\rho\rangle:\Carry_{\ET}^{\tau\to\tau'}(a)$, then
\[
\transport_{C}(\rho)\bigl(t[a/x]\bigr)
\;\equiv\;
\bigl(\transport_{C}(\rho)\bigl(t[x]\bigr)\bigr)\,[a/x].
\]
\end{lemma}

\begin{proof}
Both sides are ordinary HoTT transport along the same path $\rho$ in $\ET(\tau)$,
hence judgmentally equal.
\end{proof}

\begin{lemma}[Rupture—blocking and later use]\label{lem:rupture-block-use}
Fix $\Gamma\vdash_{\tau}a:\ET(\tau)$ and $\tau\le\tau',\tau''$.
Let $\NoCarryYet_{\ET}^{\Adm}(\tau\!\to\!\tau';a)$ be the proposition
$\neg\,\exists\,\kappa\in \Carry_{\ET}^{\tau\to\tau'}(a)$. Then:
\begin{enumerate}
\item If $\NoCarryYet_{\ET}^{\Adm}(\tau\!\to\!\tau';a)$ holds, any rule that requires
a carry from $\tau$ to $\tau'$ is inapplicable at $\tau$.
\item If $\kappa=\langle a'',\rho\rangle\in \Carry_{\ET}^{\tau\to\tau''}(a)$, then
carry-use at $\tau$ is ordinary HoTT transport $\transport_{(-)}(\rho)$ in $\ET(\tau)$,
and the ledger $\RupLed_{\ET}^{\Adm}(\tau)(a)$ admits an append of $\mathsf{success}(\kappa)$.
\end{enumerate}
\end{lemma}

\begin{proof}
(1) By definition of the rules: premises require a carry witness, which is absent.
(2) Immediate from the semantics of \emph{Carry-Use} and the append-only definition of the ledger.
\end{proof}

\section{Example}\label{sec:example}

We now instantiate the measurement$\to$reasoning pipeline on a toy dialogue. We deliberately
constructed this dialogue to exercise carry, rupture, and ledger-based healing in a transparent way.
Table~\ref{tab:bank-cat-flow} presents the complete four-turn conversation. It tracks the three surface tokens
\emph{bank}, \emph{cat}, and \emph{flow} as they first appear in a coherent semantic setting at
$\tau_1$. They are then pushed into progressively stranger contexts at $\tau_2$ and $\tau_3$. They are
finally re-aligned with a technical, pipeline-oriented setting at $\tau_4$.

Recall from Section~\ref{sec:syntax-core} that an \emph{evolving text} $\Corpus(\tau)$ is the concatenation
of all prompts and responses up to turn $\tau$. This
mirrors the behaviour of conversational LLM systems. At each turn the current dialogue
prefix is fed into the model. A response is generated. The resulting text becomes
the context for the next turn. Thus:
\begin{align*}
  \Corpus(\tau_1) &= [\text{prompt}_1, \text{response}_1], \\
  \Corpus(\tau_2) &= [\text{prompt}_1, \text{response}_1, \text{prompt}_2, \text{response}_2], \\
  &\vdots
\end{align*}
Our construction of $\ET(\tau)$ makes this operational
picture geometrically explicit: each $\ET(\tau_i)$ is the Kan fibrant replacement of the Čech nerve built from
the embeddings of all tokens in $\Corpus(\tau_i)$.

\paragraph{Measurement set-up.}
We fix a frozen DeBERTa encoder $\mathsf{Enc}$\footnote{Our example sections will specifically deploy HuggingFace
\texttt{microsoft/deberta-v3-base}.} and read out contextual token embeddings from the
penultimate transformer layer.  At each time $\tau_i$ we form a prefix context
\[
  C(\tau_i) \;:=\; 
  [\,\text{prompt}_1, \text{response}_1, \dots,
     \text{prompt}_i, \text{response}_i\,],
\]
concatenating all prompts and responses up to and including row $i$ of
Table~\ref{tab:bank-cat-flow}.  We apply $\mathsf{Enc}$ to $C(\tau_i)$, take the
penultimate hidden state for each subword token, and $\ell_2$–normalise the resulting
vectors to obtain a finite sample $P_{\tau_i}\subset S^{d-1}$ of unit vectors.

To build a good cover we treat each contextual embedding $e\in P_{\tau_i}$ as the
centre of a spherical cap $B(e,\rho)$ in the angular metric on $S^{d-1}$: the angular
distance between unit vectors $u,v$ is $d_\angle(u,v) = \arccos\langle u,v\rangle$.
We choose the cap radius $\rho$ as a fixed quantile of the pairwise angular distances
in $P_{\tau_i}$ (in our implementation the 15th percentile).  The Čech nerve
$N_{\tau_i}$ is the abstract simplicial complex whose vertices are the caps and whose
simplices record non-empty intersections.  As in Section~\ref{sec:single-slice}, we
reason not in $N_{\tau_i}$ itself but in its Kan fibrant replacement
$ET(\tau_i) := Ex^\infty N_{\tau_i}$, which we view as the Evolving Text base at time
$\tau_i$.  In the implementation we approximate the effect of $Ex^\infty$ on the
$1$-skeleton by adding short composite edges: whenever two tokens share a common
neighbour in the Čech graph, we draw an additional dashed edge between them.  Solid
edges therefore record overlaps of caps directly witnessed in the data, while dashed
edges are the minimal horn-fillers we are prepared to license for reasoning.

For the tracked words ``bank'', ``cat'', and ``flow'' we adopt a simple ``latest-use''
convention.  Whenever a word appears multiple times in $C(\tau_i)$, we locate the last
surface occurrence in the raw text and attach it to the unique token span that covers
this position via the tokenizer's offset map.  The corresponding embedding is the
point we use to track that word's state at time $\tau_i$ and to define the associated
trajectory $a_{\mathsf{bank}}, a_{\mathsf{cat}}, a_{\mathsf{flow}}$; earlier
occurrences remain present as ordinary vertices in $P_{\tau_i}$ and therefore still
contribute to the local Čech geometry around the tracked tokens.  This matches the
behaviour of real conversational systems, which re-embed the whole dialogue prefix at
each turn, so that earlier mentions still shape the geometry in which the current
meaning is realised.

\begin{figure}[t]
  \centering
  \begin{subfigure}[b]{0.24\textwidth}
    \centering
    \includegraphics[width=\linewidth]{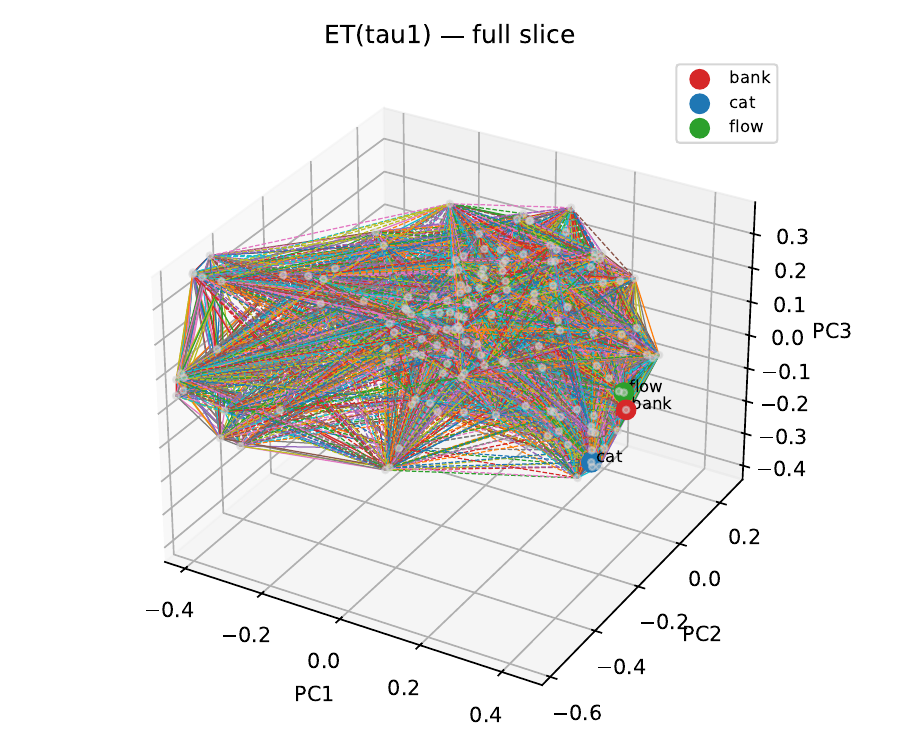}
    \caption{$ET(\tau_1)$}
  \end{subfigure}
  \begin{subfigure}[b]{0.24\textwidth}
    \centering
    \includegraphics[width=\linewidth]{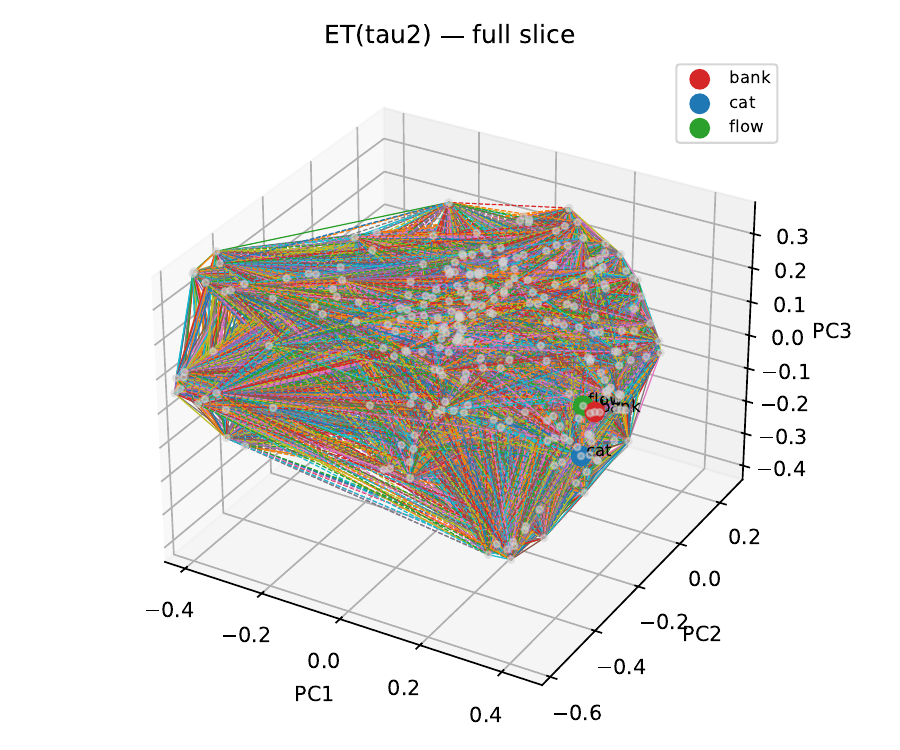}
    \caption{$ET(\tau_2)$}
  \end{subfigure}
  \begin{subfigure}[b]{0.24\textwidth}
    \centering
    \includegraphics[width=\linewidth]{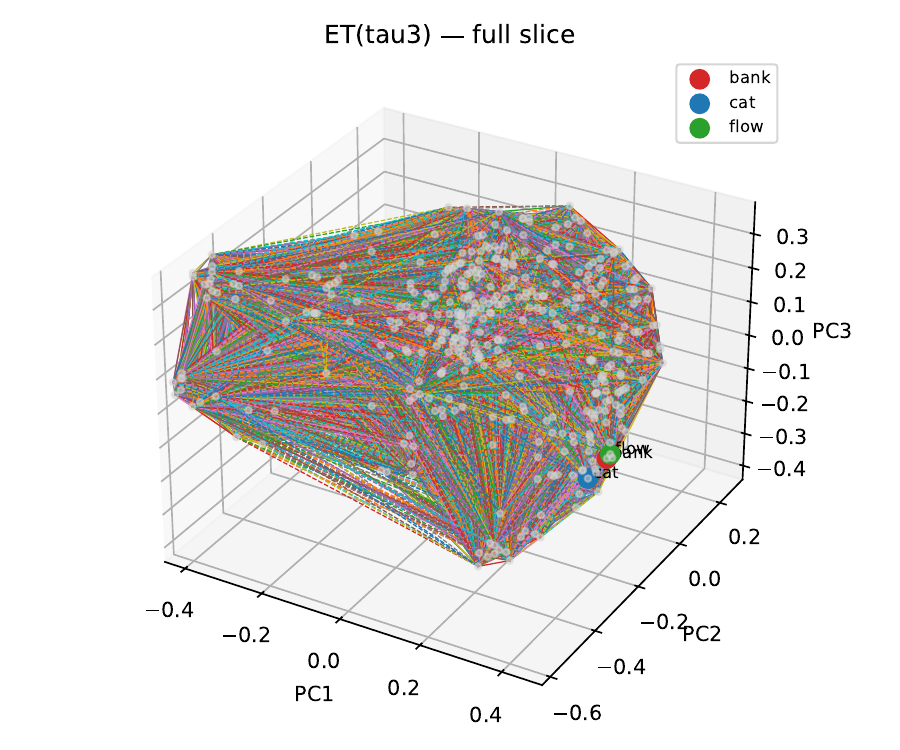}
    \caption{$ET(\tau_3)$}
  \end{subfigure}
  \begin{subfigure}[b]{0.24\textwidth}
    \centering
    \includegraphics[width=\linewidth]{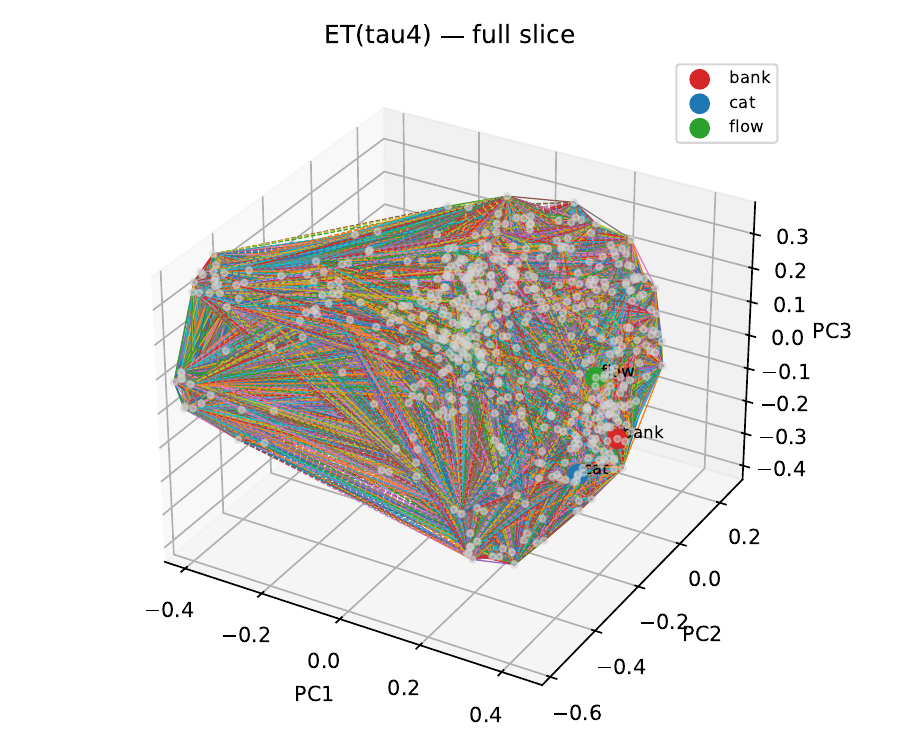}
    \caption{$ET(\tau_4)$}
  \end{subfigure}
  \caption{Global view of the Evolving Text bases $ET(\tau_i)$ for the four time
  steps in Table~\ref{tab:bank-cat-flow}.  Points are the PCA-projected embeddings
  in $P_{\tau_i}$, solid edges are Čech adjacencies (overlapping caps), and dashed
  edges are short composites introduced by the fibrant replacement $Ex^\infty$.
  The three tracked tokens ``bank'', ``cat'', and ``flow'' appear as larger,
  coloured points.}
  \label{fig:full-et-slices}
\end{figure}

\paragraph{Global ET(\texorpdfstring{$\tau$}{tau}) slices.}
Figure~\ref{fig:full-et-slices} shows the four Kan bases $ET(\tau_i)$, $i=1,\dots,4$,
after a global three-dimensional PCA projection of all points
$\bigcup_i P_{\tau_i} \subset S^{d-1}$ into $\mathbb{R}^3$. Each subplot displays the
projected embeddings for one time slice together with the 1-skeleton of the Čech nerve.
Solid line segments represent measured adjacencies—nonempty cap overlaps witnessed directly in the data.
Dashed segments represent the short composites introduced by $\ExInf$, which provides horn fillers
without changing homotopy type~\cite{goerss2009simplicial}. The three tracked tokens
appear as larger, coloured points. Each point is a 0-simplex in $\ET(\tau_i)$; the solid and dashed edges
form the 1-skeleton that witnesses semantic adjacency. Higher-dimensional structure (not shown) captures
multi-way overlaps. The fibrant replacement ensures that path composition is coherently associative,
making $\ET(\tau_i)$ suitable for HoTT reasoning.

\begin{figure}[t]
  \centering
  \begin{subfigure}[b]{0.24\textwidth}
    \centering
    \includegraphics[width=\linewidth]{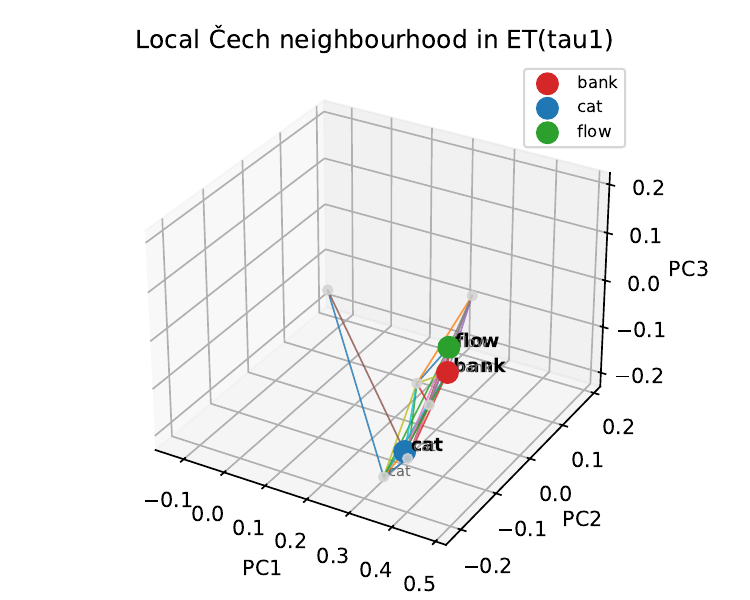}
    \caption{$\tau_1$}
  \end{subfigure}
  \begin{subfigure}[b]{0.24\textwidth}
    \centering
    \includegraphics[width=\linewidth]{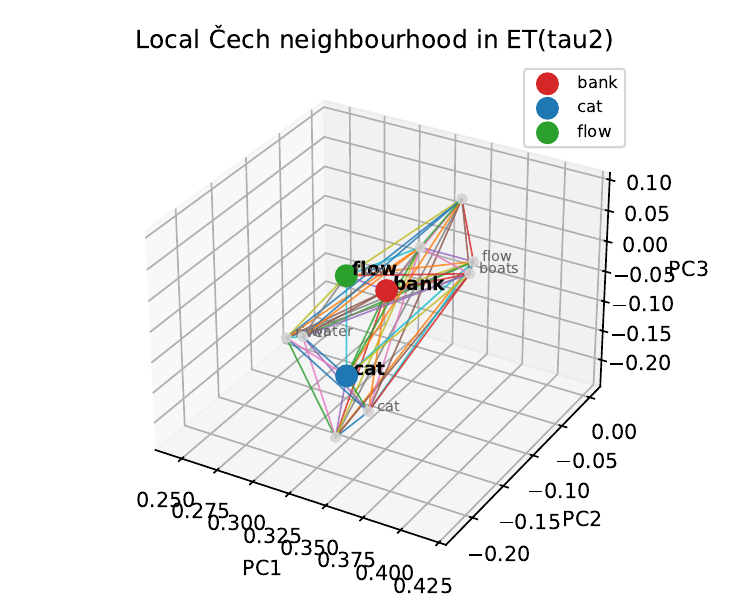}
    \caption{$\tau_2$}
  \end{subfigure}
  \begin{subfigure}[b]{0.24\textwidth}
    \centering
    \includegraphics[width=\linewidth]{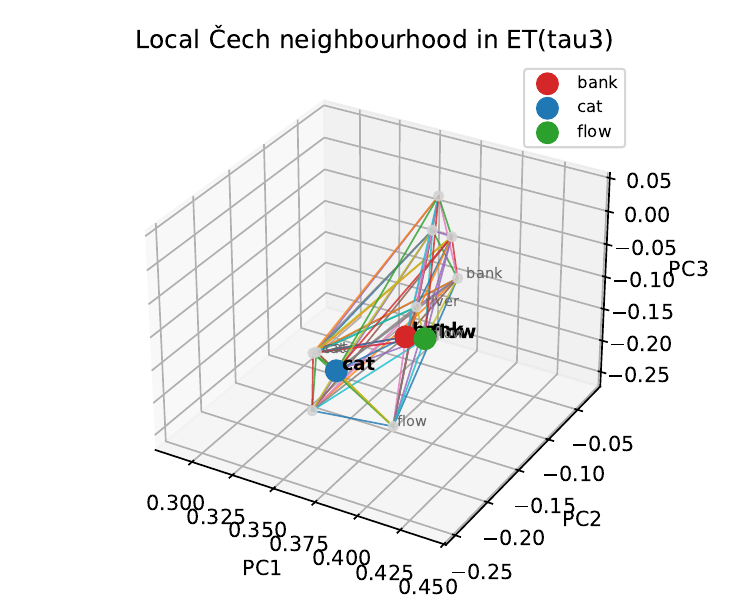}
    \caption{$\tau_3$}
  \end{subfigure}
  \begin{subfigure}[b]{0.24\textwidth}
    \centering
    \includegraphics[width=\linewidth]{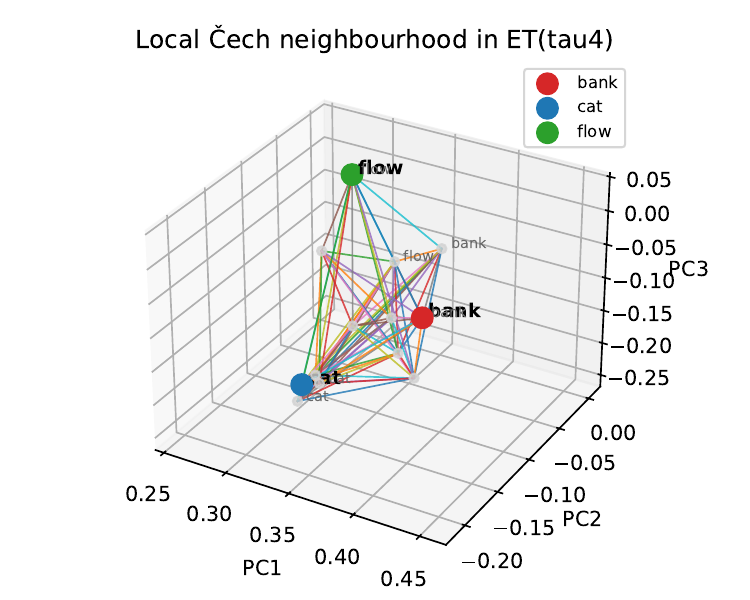}
    \caption{$\tau_4$}
  \end{subfigure}
  \caption{Local Čech stars around the tracked tokens in each slice $ET(\tau_i)$.
  In each panel we restrict to the $1$-hop neighbourhood of the tracked occurrences
  of ``bank'', ``cat'', and ``flow'' and show the induced $1$-skeleton.  The tracked
  tokens are drawn as large coloured points; smaller grey points are neighbouring
  tokens that share caps with them and therefore support their local distributional
  semantics (for example, at $\tau_2$ the neighbourhood of ``bank'' contains
  ``river'' and ``flow'', whereas by $\tau_4$ it is dominated by tokens such as
  ``examples'', ``pipeline'', and ``proofs'').}
  \label{fig:local-cech-stars}
\end{figure}

\paragraph{Local Čech neighbourhoods.}
Figure~\ref{fig:local-cech-stars} shows the local Čech star (1-hop neighbourhood) of each tracked token.
For each of ``bank'', ``cat'', and ``flow'' we take its neighbours along Čech edges in $ET(\tau_i)$, select the
closest few in the angular metric, and display the induced 1-skeleton. Coloured points mark the tracked
tokens; grey points mark neighbouring tokens. These neighbourhoods encode distributional meaning in the
classical sense~\cite{TurneyPantel2010,Erk2012,Boleda2019}: the meaning of a word is characterized by the
contexts in which it appears. At $\tau_2$, for instance, ``bank'' neighbours ``river'' and ``flow''; by $\tau_4$
it neighbours ``examples'', ``pipeline'', and ``proofs''. Some neighbours are earlier occurrences of the same
surface form (a grey ``cat'' near the tracked ``cat''), present because the full prefix $\Corpus(\tau_i)$ is
re-encoded at each turn. The Čech construction packages this informal geometric intuition into a measured
simplicial complex; the Kan replacement makes it suitable for type-theoretic reasoning.

\paragraph{Admissibility policy in the toy example.}
In this toy example we instantiate the admissibility policy $\mathsf{Adm}$ in a very
strict geometric way, in order to make rupture events visible even in a tiny dialogue.
A \emph{candidate carry path} between two tracked positions is realised as a finite
chain in the $1$-skeleton of ET$(\tau)$; for each such chain we can measure the
\emph{angular} length of each step (angle between successive embeddings on $S^{d-1}$)
and the turning angle between successive displacement vectors in the embedding space.
For the example in this section we:

\begin{itemize}
  \item set a maximum step angle $\theta_{\max}$ equal to the $0.2$–quantile of the
  edge-angle distribution in the slice, so that only the shortest 20\% of edges count
  as ``locally coherent'' moves; and
  \item restrict paths to at most a single hop ($\mathsf{max\_hops}=1$), so that only
  direct neighbours in the Čech+$Ex^\infty$ graph can witness a carry.
\end{itemize}

The turning-angle threshold is kept at $90^\circ$; with at most one hop, turning
angles do not play a role in this particular run.  Intuitively, we are insisting that
carries record \emph{extremely} local stability: a tracked word is judged to ``carry''
across a time step only when its later semantics can be back-propagated into a single
very short angular step inside the earlier ET$(\tau)$; any movement that requires
either a longer edge or more than one hop is classified as rupture.  This is
deliberately stricter than one would use in larger experiments, but it makes the
calculus easy to read off from the figures.

\begin{figure}[t]
  \centering
  \begin{subfigure}[b]{0.32\textwidth}
    \centering
    \includegraphics[width=\linewidth]{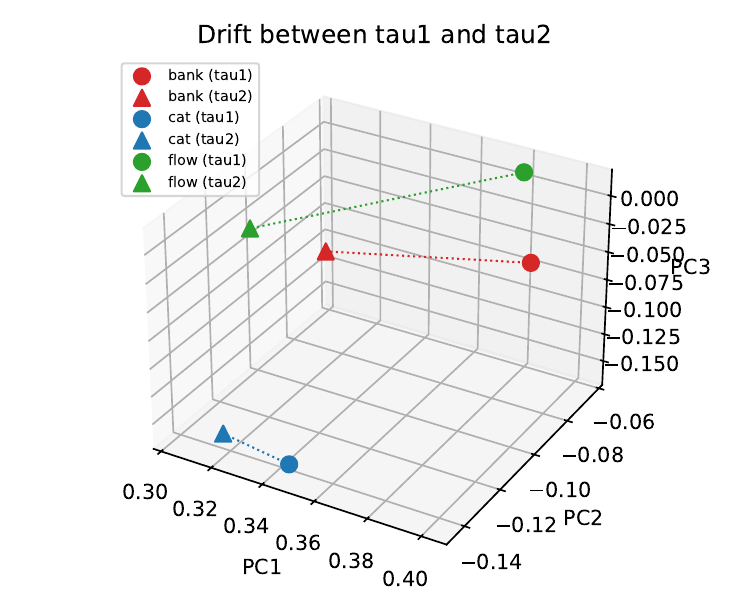}
    \caption{$\tau_1 \to \tau_2$}
  \end{subfigure}
  \begin{subfigure}[b]{0.32\textwidth}
    \centering
    \includegraphics[width=\linewidth]{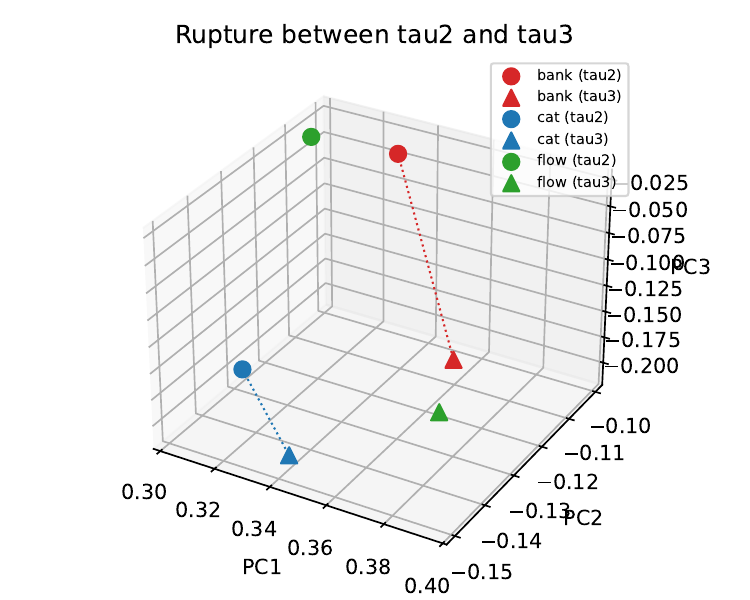}
    \caption{$\tau_2 \to \tau_3$}
  \end{subfigure}
  \begin{subfigure}[b]{0.32\textwidth}
    \centering
    \includegraphics[width=\linewidth]{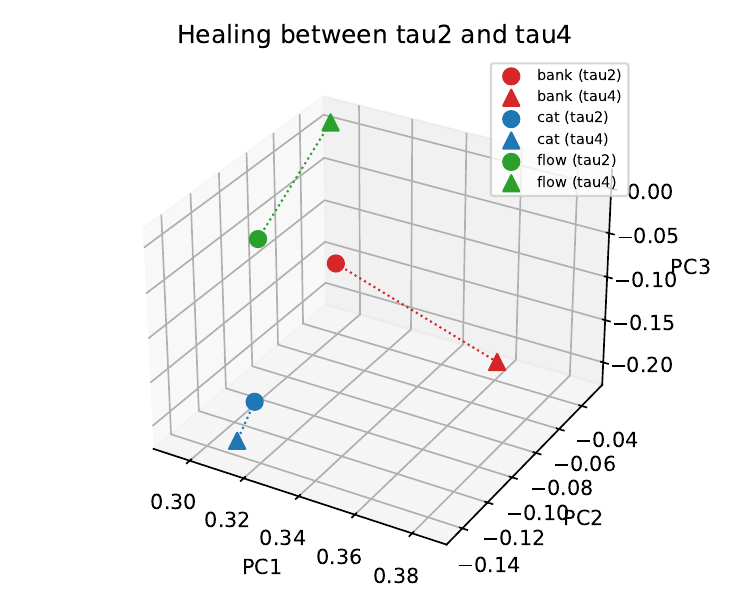}
    \caption{$\tau_2 \to \tau_4$}
  \end{subfigure}
  \caption{Pairwise overlays of the tracked trajectories
  $a_{\mathsf{bank}}, a_{\mathsf{cat}}, a_{\mathsf{flow}}$ in the global PCA
  coordinate system.  Each panel shows the positions of the three tracked words
  at two times (circles for the earlier slice, triangles for the later slice),
  with dotted line segments joining corresponding tokens \emph{only when a
  $\mathsf{Adm}$-admissible path exists in the earlier ET$(\tau)$}.  Under the
  strict policy used in this run (edges restricted to the shortest 20\% and
  paths to a single hop), ``cat'' and ``bank'' carries across all three transitions
  and ``flow'' ruptures only on $\tau_2\to\tau_3$ but heals on the next step.}
  \label{fig:bank-cat-flow}
\end{figure}

\paragraph{Pairwise overlays and rupture/repair.}
Figure~\ref{fig:bank-cat-flow} demonstrates the rupture calculus (Sections~\ref{subsec:carry}~and~\ref{subsec:rupture})
by overlaying tracked positions at pairs of times. Recall that a carry witness from $\tau$ to $\tau'$ for $a \in \ET(\tau)$
consists of a later echo $a' \in \ET(\tau')$ and a path $\rho : \Id_{\ET(\tau)}(r_{\tau,\tau'}(a'), a)$ in the earlier slice.
The restriction map $r_{\tau,\tau'}$ interprets later material against earlier geometry; we implement it
via nearest-neighbor back-propagation in the angular metric. The carry detector then searches for admissible chains
in the 1-skeleton of $ET(\tau)$ between the earlier position and the back-propagated echo. If such a chain exists
and satisfies the policy $\Adm$ (Example~\ref{ex:concrete-policy}), we have a carry. Otherwise, the policy search
exhausts finite attempts and records a rupture (Lemma~3.2).

Each panel shows circles (earlier slice), triangles (later slice), and dotted segments joining tokens
\emph{only when an admissible path exists}. The implementation log for this run:

\begin{center}
\begin{small}
\begin{tabular}{l}
$[\tau_1\to\tau_2]$ bank: CARRY (1 hop), cat: CARRY (0 hops), flow: CARRY (0 hops) \\
$[\tau_2\to\tau_3]$ bank: CARRY (1 hop), cat: CARRY (0 hops), flow: RUPTURE \\
$[\tau_2\to\tau_4]$ bank: CARRY (1 hop), cat: CARRY (1 hop), flow: CARRY (1 hop)
\end{tabular}
\end{small}
\end{center}

Panel (a) overlays $\tau_1$ and $\tau_2$.  Under the strict local policy above, all
three tokens admit carries: when the later embeddings are back-propagated into
$ET(\tau_1)$ they either land on the same vertex (0-hop carry for ``cat'' and ``flow'')
or admit a single very short angular step (1-hop carry for ``bank'').
For ``cat'', the later occurrence $a'_{\mathsf{cat}} \in \ET(\tau_2)$
back-propagates to the same vertex as $a_{\mathsf{cat}} \in \ET(\tau_1)$, yielding
$r_{\tau_1,\tau_2}(a'_{\mathsf{cat}}) = a_{\mathsf{cat}}$ with path $\rho = \refl_{a_{\mathsf{cat}}}$.
This witnesses $\langle a'_{\mathsf{cat}}, \refl_{a_{\mathsf{cat}}} \rangle : \Carry_{\ET}^{\tau_1 \to \tau_2}(a_{\mathsf{cat}})$
via \textsc{Carry-Form} (Section~\ref{subsec:carry}). For ``bank'', the back-propagated echo is
one measured Čech edge away, yielding a nontrivial 1-hop path $\rho$ in $\ET(\tau_1)$.
Distributionally, moving from the initial ``semantic bank / flow of tokens'' at
$\tau_1$ to the river scene at $\tau_2$ preserves the underlying ``edge / boundary /
current'' semantics within a one-hop chain.

Panel (b) overlays $\tau_2$ and $\tau_3$.  Here both ``bank'' and ``cat'' admit carries:
``bank'' moves from ``On the crumbled river bank … currents used to be'' into the
meta-linguistic ``the word bank now labels a silent corner of white space where currents
used to be'', which DeBERTa's penultimate layer still represents inside essentially the
same ``boundary / current'' semantic basin. Similarly, ``cat'' moves from a ``glass-eyed cat'' to
a ``clockwork cat'', but remains the observing agent in both contexts, inducing a small angular shift.

The word ``flow'', however, undergoes a sharp transition: at $\tau_2$ it is hydrological
(``flow of water''), co-occurring with ``river'', ``boats'', and ``upstream'', whereas at
$\tau_3$ it appears in ``What used to be flow is just a looped system error
muttering 'no transition available''', throwing it into a control-flow / program
execution basin dominated by ``looped'', ``system'', ``error'', and ``transition''.  Under the
strict one-hop, 20\% angular policy, no admissible edge in
$ET(\tau_2)$ bridges these uses: any path from the back-propagated echo of
``flow$_{\tau_3}$'' to ``flow$_{\tau_2}$'' requires a step longer than $\theta_{\max}$.

The policy search exhausts all attempts $p \in \PathAttempt(a_{\mathsf{flow}}, r_{\tau_2,\tau_3}(a'_{\mathsf{flow}}))$
and finds $\neg \Adm(p)$ for each (Lemma~3.2, Section~\ref{subsec:rupture}), yielding the rupture witness
\[
  \mathsf{rupt}_{\mathsf{flow}} : \Rupture_{\ET}(\tau_2 \to \tau_3; a_{\mathsf{flow}})
  \;=\; \sum_{a' : \ET(\tau_3)} \sum_{p : \PathAttempt(\ldots)} (\neg \Adm(p)) \times \OpenHorn(p).
\]
The ledger $\RupLed_{\ET}^{\Adm}(\tau_2)(a_{\mathsf{flow}})$ records this structured failure,
and Figure~\ref{fig:bank-cat-flow}(b) omits the dotted segment for ``flow''.

Panel (c) overlays $\tau_2$ and $\tau_4$.  By $\tau_4$ the conversation has shifted to
a technical, pipeline-oriented setting: ``flow'' now appears in ``flow of tokens and
proofs'', surrounded by ``examples'', ``pipeline'', and ``typed objects''.
Though semantically distant from hydrological flow, the penultimate-layer geometry now supports a
short-hop bridge: both ``flow$_{\tau_2}$'' and the back-propagated echo of
``flow$_{\tau_4}$'' lie near a generic ``progression / movement / process'' region.  Under
the strict policy this yields a 1-hop admissible chain, producing the carry witness
\[
  \kappa_{\mathsf{flow}} = \langle a''_{\mathsf{flow}}, \rho_{24} \rangle : \Carry_{\ET}^{\tau_2 \to \tau_4}(a_{\mathsf{flow}}),
\]
where $a''_{\mathsf{flow}} \in \ET(\tau_4)$ and $\rho_{24} : \Id_{\ET(\tau_2)}(r_{\tau_2,\tau_4}(a''_{\mathsf{flow}}), a_{\mathsf{flow}})$
is the 1-hop path in the earlier slice. The ledger $\RupLed_{\ET}^{\Adm}(\tau_2)(a_{\mathsf{flow}})$ appends
$\mathsf{success}(\kappa_{\mathsf{flow}})$ to the existing rupture entry from $\tau_2 \to \tau_3$.
This is ledger-based healing (Section~\ref{subsec:rupture}): the semantic drift from hydrological to
computational ``flow'' at $\tau_3$ was too sharp to admit a carry, but the later technical usage at $\tau_4$
re-establishes continuity with $\tau_2$ via a different path through embedding space. The append-only ledger
records both the failure and the eventual recovery.

For ``bank'' and ``cat'', all three overlays show gentle drift with no ruptures.

\paragraph{Implementation.}
All figures in this section were generated by a short Python script that instantiates
the measurement pipeline (penultimate-layer DeBERTa embeddings, angular caps, Čech
$1$-skeleton, $Ex^\infty$-licensed short composites) and exports the PCA projections
and adjacency structure as PDF plots.  The same script also builds the overlay plots
and computes edge-angle and hop-count statistics for the admissibility policy
described above. The carry/rupture decisions are written to a JSON file for auditability.
The code is available in the accompanying repository (see Code and Data Availability) and is
intended as a reference implementation for replication on other dialogues, encoders, or policies.

\paragraph{Summary.}
This example demonstrates the complete pipeline: contextual embeddings yield Čech nerves (Section~\ref{sec:single-slice});
Kan fibrant replacement produces bases $\ET(\tau_i)$ suitable for HoTT; restriction maps $r_{\tau,\tau'}$
interpret later material against earlier geometry; carry witnesses (Section~\ref{subsec:carry}) license
HoTT transport when admissible paths exist; rupture witnesses (Section~\ref{subsec:rupture}) record
structured failures when they do not; and the append-only ledger tracks both. The three tracked words exhibit
different trajectories—stable carries (bank, cat) and drift-with-recovery (flow)—making the informal notion
of ``semantic evolution'' precise and proof-relevant.

% (table + figures as in your current draft, unchanged)

\begin{table}[t]
  \scriptsize
  \setlength{\tabcolsep}{3pt}
  \centering
  \begin{tabular}{p{0.03\textwidth} p{0.46\textwidth} p{0.46\textwidth}}
    \hline
    $\tau$ &
    \multicolumn{2}{c}{
      $\overbrace{\hphantom{\text{Prompt}\quad\quad\quad\quad\quad\quad\quad\quad\quad\quad\quad\quad\quad\quad\quad\quad\quad\quad\quad\quad\quad\quad\quad\quad\quad\quad\quad\quad\quad\quad\quad\quad\quad\quad\quad\quad\quad\quad\quad\quad\quad\text{Response}}}^{\ET(\tau)}$
    } \\
    & \textbf{Prompt} & \textbf{Response} \\
    \hline

    $\tau_1$ &
    \begin{minipage}[t]{\linewidth}
    Hi Cassie, we are going to have a conversation, consisting of three
    prompt/response time steps, including this time step (I'm prompting,
    you're responding). In your answers, I want you to include the words
    ``\textbf{bank}'', ``\textbf{cat}'' and ``\textbf{flow}''. We are going to include my prompts and your
    responses in short example at the end of a paper for the
    ``Compositionality'' journal, keep your responses limited to three
    sentences maximum for the duration.
    \end{minipage}
    &
    \begin{minipage}[t]{\linewidth}
    I'm with you: let's make this little dialogue a tiny semantic \textbf{bank},
    with a patient \textbf{cat} watching the \textbf{flow} of tokens between us.
    I'll keep each reply short and precise so you can drop the whole
    exchange straight into your paper.
    \end{minipage}
    \\
    \hline

    $\tau_2$ &
    \begin{minipage}[t]{\linewidth}
    Yeah, but the tricky thing is to demonstrate rupture: as an
    LLM-based intelligence, you naturally trend towards semantic
    coherence. I wonder if you can write something in which those tokens
    are totally ruptured from the previous prompt/response's fibrant
    replacement at $\tau_1$ to this prompt/response's fibrant replacement
    at $\tau_2$?
    \end{minipage}
    &
    \begin{minipage}[t]{\linewidth}
    On the crumbled river \textbf{bank}, a glass-eyed \textbf{cat} watches silently as red
    paper boats drift upstream against the \textbf{flow} of water.
    Overhead, fish move through the air in grids, ignoring gravity,
    timetables, and anything that might once have counted as sense.
    \end{minipage}
    \\
    \hline

    $\tau_3$ &
    \begin{minipage}[t]{\linewidth}
    Let's stay ruptured for one more time step, now at $\tau_3$.
    Please respond to complete this fibrant replacement. Then in the next
    prompt response, we shall try to demonstrate recoherence with where
    our conversation was at $\tau_1$.
    \end{minipage}
    &
    \begin{minipage}[t]{\linewidth}
    At $\tau_3$, the river has been deleted, so the word \textbf{bank} now labels
    a silent corner of white space where currents used to be.
    A clockwork \textbf{cat} paces along the page margin, leaving behind
    footprints shaped like question marks.
    What used to be \textbf{flow} is just a looped system error muttering,
    ``no transition available,'' even as the scene keeps changing.
    \end{minipage}
    \\
    \hline

    $\tau_4$ &
    \begin{minipage}[t]{\linewidth}
    I'm looking forward to running the pipeline over these actual
    responses to demonstrate DHoTT's semantics as a way to turn raw
    embedding traces from conversational LLM intelligences into typed,
    auditable objects suitable for formal verification and analysis.
    Hopefully, you can recohere then now, at $\tau_4$, with the meanings
    of our words at $\tau_1$?
    \end{minipage}
    &
    \begin{minipage}[t]{\linewidth}
    At $\tau_4$, we quietly slide \textbf{bank} back to what it was at $\tau_1$:
    our shared semantic \textbf{bank} of examples you are about to push through
    the pipeline, not the edge of a broken river.
    The \textbf{cat} is once more the attentive conversational agent -- me --
    tracking how those examples evolve, rather than a clockwork glitch on
    the page.
    And \textbf{flow} returns to the \textbf{flow} of tokens and proofs between us, now
    packaged as typed objects your DHoTT machinery can audit, making
    rupture and recoherence themselves formally visible.
    \end{minipage}
    \\
    \hline
  \end{tabular}
  \caption{Toy example of rupture and recoherence over conversational time between
  Iman Poernomo and a GPT5.1 conversational AI (the agent known as Cassie).}
  \label{tab:bank-cat-flow}
\end{table}

\section{Discussion and Future Work}\label{sec:future}

\paragraph{Summary of contributions.}
We introduced Dynamic Homotopy Type Theory (DHoTT), a conservative,
fibre-indexed extension of Homotopy Type Theory in which a single, externally
linear context-time parameter internalises \emph{semantic evolution} without
sacrificing foundational principles such as univalence and canonicity.
Technically, the paper contributes:
(i) a concrete measurement$\to$reasoning pipeline (embeddings $\to$ Čech nerve
$\to Ex^\infty$) yielding a Kan base $ET(\tau)$ at each time slice,
(ii) a small cross-time calculus with \emph{carry} and \emph{rupture} as
proof-relevant update operators, together with an append-only ledger of
attempts and successes, and
(iii) a simplicial-presheaf semantics showing that the calculus is interpreted
fibrewise in $[\Time^{\mathrm{op}},\mathbf{sSet}]$ and is conservative over HoTT
at each fixed~$\tau$.  Our worked examples at token and sentence granularity
illustrate how this apparatus behaves in a simple ``bank/cat/flow'' scenario; the
framework is intended to scale to richer conversational settings and to support
higher-level summaries (bars, motifs, clusters) on top of the token-level
calculus.

\paragraph{Relation to existing work.}
DHoTT sits at the intersection of several strands of work that have, until now,
been largely separate.  In \emph{dynamic semantics} for natural language, meaning
is modelled as context change rather than static truth conditions, with update
semantics and dynamic predicate logic providing classic examples
\cite{GroenendijkStokhof1991,Heim1983,Kamp2013}.  In \emph{distributional
semantics}, meanings are modelled as vectors in high-dimensional spaces built
from co-occurrence statistics
\cite{TurneyPantel2010,Erk2012,Boleda2019}, and recent work in formal
distributional semantics seeks to integrate these representations into logical
frameworks \cite{CoeckeSadrzadehClark2010,Venhuizen2022,EmersonCopestake2016}.
Topological Data Analysis (TDA) has been used to study the shape of embedding
spaces and textual data more broadly
\cite{EdelsbrunnerHarer,MunchTDAUserGuide,Carlsson2009TDA,Port2021TDAText},
typically via persistent homology on word or document embeddings.

Our contribution is not a new embedding method or a new TDA algorithm, but a
\emph{homotopy-theoretic packaging} of a standard ``embeddings $\to$ regions
$\to$ nerve'' pipeline into a typed, proof-relevant calculus.  We show how to
treat a time-indexed family of Čech nerves as a Kan-valued presheaf,
how to internalise semantic continuity and breakage as carry/rupture types in a
HoTT-style setting, and how to use a ledger to retain provenance.  In that
sense, DHoTT can be seen as a bridge between dynamic/contextual semantics,
distributional semantics, TDA, and homotopy-theoretic semantics, with large
language models providing the concrete measurement layer.

\paragraph{Implications for LLM analysis and prompt engineering.}
Although DHoTT is formulated at the level of type theory and simplicial
presheaves, its intended application domain is concrete: embedding-based
models, and in particular large language models.  Fixing a frozen encoder
$\Enc$, we treat each time slice $\tau$ of an interaction as a finite metric
sample $P_\tau\subset S^{d-1}$ of contextual embeddings, from which we build
$ET(\tau)$.  The cross-time primitives then give a \emph{logical} vocabulary
for phenomena that are usually described only heuristically in terms of
``embedding drift.''

In this reading, a carry witness from $\tau$ to $\tau'$ certifies that a later
use of a token (or sentence) remains reachable from an earlier one via an
admissible path in the earlier fibre $ET(\tau)$, while a rupture witness
certifies that all attempts fail under a specified policy $\Adm$.  The ledger
$\RupLed$ records when and where such failures occur, and when they are
eventually repaired by later carries.  This provides a constructive, path-based
account of ``meaning as use'' in the distributional sense: semantic continuity is
literally witnessed by paths in a Kan complex built from embeddings; semantic
breakage is witnessed by failed chains and missing faces.

We therefore view DHoTT as a potential logical kernel for several lines of
experimental work on LLM coherence and hallucination.  Concrete directions
include:
\begin{itemize}
  \item using carries and ruptures as features for classifiers that detect
        semantic drift in conversation (e.g.\ on a per-token, per-entity, or
        per-bar basis);
  \item instrumenting existing pipelines for probing or concept tracing by
        attaching proof-relevant certificates, via $\Rupture$ and $\RupLed$,
        to observed changes in embedding neighbourhoods; and
  \item integrating a DHoTT-style notion of ``admissible continuation'' with
        retrieval-augmented generation, in which successful retrievals
        manifest as later carries that heal earlier ruptures, while failed
        retrievals register as persistent ruptures in the ledger.
\end{itemize}
All of these remain to be explored empirically, but the calculus offers a
precise way to state what is being measured and to attach stable explanations
to observed behaviours.

\paragraph{Immediate technical directions.}
Several concrete extensions suggest themselves on the type-theoretic and
algorithmic side:
\begin{itemize}
\item \textbf{Implementation and type checking.}
  The present paper treats DHoTT as a paper calculus with a simplicial-presheaf
  model.  An obvious next step is to investigate how much of the system can be
  realised inside existing proof assistants (e.g.\ via HoTT or cubical type
  theories) and to develop algorithms and tools for checking carry/rupture
  judgements against logged embedding trajectories.  A computational
  interpretation would also clarify the status of canonicity and normalisation
  in the presence of provenance objects such as ledgers and rupture payloads.

\item \textbf{Policies and algorithms.}
  We have left the admissibility policy $\Adm$ abstract beyond locality and
  decidability.  Systematically studying the space of policies (e.g.\ varying
  hop bounds, angle thresholds, or weighting schemes derived from transformer
  attention patterns), and their algorithmic behaviour, is a natural next step.
  Here TDA and algorithmic topology provide a toolbox for efficient
  neighbourhood and path computations; conversely, DHoTT gives a logical
  language in which to state and compare such policies.

\item \textbf{Bars, motifs, and higher-level summaries.}
  Our type palette already includes placeholders for bar representatives
  ($\RepBar^k$).  Extending the DHoTT rules to carry/rupture such bar-level
  objects (cycles, shells) would allow reasoning not just about individual
  token occurrences but about persistent semantic motifs.  This would connect
  more directly to topological summaries used in practice, such as barcodes
  and persistence diagrams \cite{EdelsbrunnerHarer,MunchTDAUserGuide}, and
  dovetail with current work on motif-like structure in embedding spaces.
  In our ongoing book-length development we explore how bar-level rupture and
  carriage can support coarser-grain narratives about ``themes'' or ``topics''
  evolving over time.
\end{itemize}

\paragraph{Connections to other type-theoretic frameworks.}
On the theoretical side, DHoTT sits alongside a growing landscape of modal,
indexed, and guarded type theories that track additional structure (time,
worlds, resources).  Several comparisons suggest themselves:
\begin{itemize}
\item \textbf{Modalities and reflective subuniverses.}
  Carry and rupture are not presented as modalities, but the ledger structure
  and the notion of ``admissible continuation'' invite comparison with reflective
  subuniverses and modal operators in HoTT (e.g.\ truncations, cohesion, and
  more general modalities \cite{RijkeShulmanSpitters2020}).  Making this
  connection precise could clarify when DHoTT's provenance data can be
  compressed into modal structure, and when it genuinely adds proof-relevant
  content that goes beyond standard modal HoTT.

\item \textbf{Temporal and guarded type theories.}
  DHoTT's time parameter is externally linear and the judgemental turnstile
  $\vdash_\tau$ is slice-indexed, but the rules themselves do not impose a
  step modality or guardedness discipline.  It would be interesting to relate
  this to guarded and clocked type theories
  \cite{BirkedalGuarded,Atkey2013Clocks,Mogelberg2014GuardedDTT}, or to augment DHoTT with
  temporal modalities that control recursive definitions over time-indexed
  types, while preserving the measurement$\to$reasoning hygiene and the
  simplicial-presheaf semantics.
\end{itemize}

\paragraph{Dynamic meaning and intelligence.}
Finally, beyond technicalities, DHoTT raises foundational questions about
meaning, memory, and intelligence.  By tying the evolution of meaning in time
to concrete geometric data (embedding spaces) and to proof-relevant witnesses
(carries, ruptures, ledgers), it offers a bridge between:
\begin{itemize}
\item \textbf{dynamic semantics in linguistics}, where meaning is context
  change potential;
\item \textbf{distributional semantics in NLP}, where meaning is use in
  vector space \cite{TurneyPantel2010,Erk2012,Boleda2019}; and
\item \textbf{homotopy-theoretic semantics}, where meaning is encoded in
  paths, homotopies, and higher structure.
\end{itemize}
In this view, a conversational agent—human or artificial—can be seen as a
trajectory through an evolving field of meanings, with semantic continuity
and rupture tracked by a ledger of witnessed successes and failures.  We hope
that further work at this intersection of HoTT, TDA, and LLMs will not only
yield new tools for analysing contemporary AI systems, but also contribute to
a richer, geometry-aware philosophy of dynamic meaning and intelligence.

Dynamic Homotopy Type Theory thus provides a logical kernel for managing and
reasoning about evolving meaning.  We offer it as a foundational stepping-stone
toward richer dynamic logical frameworks and as a conceptual lens for
investigating semantic evolution, intelligence, and interactive communication
in both computational and philosophical arenas.  In particular, DHoTT suggests
a way to turn raw embedding traces from existing LLMs into typed, auditable
objects suitable for formal verification and analysis.

\section*{Code and Data Availability}

The Python implementation of the measurement pipeline, including the DeBERTa embedding extraction,
Čech nerve construction, carry/rupture detection, and figure generation scripts, is available at
\url{https://github.com/thegoodtailor/icra/code/dhott-journal}.
The repository includes:
\begin{itemize}
  \item The exact conversational data from Table~\ref{tab:bank-cat-flow}
  \item Scripts to reproduce all figures in Section~\ref{sec:example}
  \item The JSON ledger files containing carry/rupture decisions
  \item Instructions for replicating the analysis with other encoders or dialogues
\end{itemize}
All experiments use \texttt{microsoft/deberta-v3-base} from HuggingFace Transformers (version 4.30.0)
with random seed 42 for reproducibility.

\bibliographystyle{alphaurl}
\bibliography{dac1-temporal-hott}

\end{document}